\documentclass[aps,manuscript,showpacs,showkeys,superscriptaddress,nofootinbib]{revtex4-1}
\RequirePackage{graphicx}
\usepackage{graphicx}% Include figure files
\usepackage{epsfig}  % Include figure files
\usepackage{epsf}    % Include figure files
\usepackage{dcolumn}% Align table columns on decimal point
\usepackage{bm}% bold math
\usepackage{dcolumn}% Align table columns on decimal point
\usepackage{textcomp}% Align table columns on decimal point
\usepackage[tbtags]{amsmath}
\usepackage{amsfonts}
\usepackage{float}
\usepackage{subfig}
\usepackage[]{hyperref}
  \hypersetup{
  unicode=false,          % non-Latin characters in Acrobat?s bookmarks
  pdftoolbar=true,        % show Acrobat?s toolbar?
  pdfmenubar=true,        % show Acrobat?s menu?
  pdffitwindow=true,     % window fit to page when opened
  pdfstartview={FitH},    % fits the width of the page to the window
  pdfsubject={Getting the best out of T2K and NOvA},   % subject of the document
  pdfnewwindow=true,      % links in new window
  pdfcreator={RevTeX},
  colorlinks=true,       % false: boxed links; true: colored links
  linkcolor=red,          % color of internal links
  citecolor=blue,        % color of links to bibliography
  urlcolor=blue,           % color of external links
  }
\usepackage{hypcap}

\newcommand{\beq}{\begin{equation}}
\newcommand{\eeq}{\end{equation}}
\newcommand{\beqa}{\begin{eqnarray}}
\newcommand{\eeqa}{\end{eqnarray}}

\newcommand{\tx}{{\theta_{12}}}
\newcommand{\ty}{{\theta_{13}}}
\newcommand{\tz}{{\theta_{23}}}

\newcommand{\dl}{{\Delta_{31}}}
\newcommand{\ds}{{\Delta_{21}}}

\newcommand{\ahat}{\hat{A}}
\newcommand{\dhat}{\hat{\Delta}}

\newcommand{\dcp}{\delta_{\mathrm{CP}}}
\newcommand{\nova}{NO$\nu$A~}

\newcommand{\pme}{P_{\mu e}}

\newcommand{\pmebar}{P_{\bar{\mu} \bar{e}}}

\newcommand{\dchsq}{\Delta\chi^2}

\begin{document}

\title{Effect of non-unitary mixing on the mass hierarchy and CP violation determination at the Protvino to ORCA experiment%\thanksref{t1}
}
%\subtitle{Do you have a subtitle?\\ If so, write it here}

%\titlerunning{Short form of title}        % if too long for running head

\author{Daljeet Kaur}
\email[Email Address: ]{daljeet.kaur97@gmail.com}
\affiliation{SGTB Khalsa College, University of Delhi, New Delhi-110007, India}
\author{Nafis Rezwan Khan Chowdhury}
\email[Email Address: ]{nafis.chowdhury@ific.uv.es}
\email[Email Address:]{nafis.chowdhury@utah.edu}
\affiliation{IFIC \textendash \ Instituto de F\'{i}sica Corpuscular (Univ. de Val\'{e}ncia \textendash \ CSIC) \\
Catedr\'{a}tico Jos\'{e} Beltr\'{a}n, 2, E-46980 Paterna, Espa\~{n}a}
\affiliation{Department of Physics and Astronomy, University of Utah, \\
115 S. 1400 E., Salt Lake City, Utah 84112}

\author{Ushak Rahaman}
\email[Email Address: ]{ushakr@uj.ac.za}
\email[Email Address: ]{ushak.rahaman@ific.uv.es}
\affiliation{Centre for Astro-Particle Physics (CAPP) and Department of Physics, University of Johannesburg, PO Box 524, Auckland Park 2006, South Africa}
\affiliation{IFIC \textendash \ Instituto de F\'{i}sica Corpuscular (Univ. de Val\'{e}ncia \textendash \ CSIC) \\
Catedr\'{a}tico Jos\'{e} Beltr\'{a}n, 2, E-46980 Paterna, Espa\~{n}a}

\begin{abstract}
In this paper, we have estimated the neutrino mass ordering and the CP violation sensitivity of the proposed Protvino to ORCA (P2O) experiment after 6 years of data-taking. Both unitary and non-unitary $3\times 3$ neutrino mass mixing have been considered in the simulations. A forecast analysis deriving possible future constraints on non-unitary parameters at P2O have been performed.
\end{abstract}
% \PACS{PACS code1 \and PACS code2 \and more}
% \subclass{MSC code1 \and MSC code2 \and more}

\maketitle

\section{Introduction}
\begin{table}
  \begin{center}
\begin{tabular}{|c|c|c|}
%  \showrowcolors
  \hline
  Parameters & NH \phantom{foo}&
  IH\\
  \hline %\hiderowcolors
  $\tx/^o$ & $33.44^{+0.77}_{-0.74}$ & $33.45^{+0.78}_{-0.75}$\\
  \hline
  $\tz/^o$ & $49.2^{+0.9}_{-1.2}$ & $49.3^{+0.9}_{-1.1}$\\
  \hline
  $\ty/^o$ & $8.57^{+0.12}_{-0.12}$ & $8.60^{+0.12}_{-0.12}$\\
  \hline
  $\dcp/^o$  & $197^{+24}_{-24}$ & $282^{+26}_{-30}$ \\
  \hline
 $ \frac{\ds}{10^{-5}\, {\rm eV}^2}$ & $7.42^{+0.21}_{-0.20}$ & $7.42^{+0.21}_{-0.20}$ \\
 \hline
  $ \frac{\Delta_{3l}}{10^{-3}\, {\rm eV}^2}$ & $2.517^{+0.026}_{-0.028}$ & $-2.498^{+0.028}_{-0.028}$ \\
 \hline
\end{tabular}
\end{center}
 \caption{Global best-fit values of neutrino oscillation parameters \cite{nufit, Esteban:2020cvm}. $\Delta_{3l}=\dl>0$ ($\Delta_{32}<0$) for NH (IH).}
  \label{bestfit}
\end{table}
Neutrino oscillation is parameterised by the the mixing of three mass eigen states 
into three flavour states by the following mixing scheme:
\begin{equation}
 \left[ {\begin{array}{c}
   \nu_e  \\
   \nu_\mu \\
   \nu_\tau
  \end{array} } \right]= U  \left[ {\begin{array}{c}
   \nu_1  \\
   \nu_2 \\
   \nu_3
  \end{array} } \right],
\end{equation}
where the vector on the left (right) hand side denotes the neutrino flavour (mass) eigen states. $U$ is the 
$3\times 3$ unitary PMNS mixing matrix parameterised as
\begin{equation}
U= \left[
\begin{array}{ccc}
c_{13}c_{12} & s_{12}c_{13} & s_{13}e^{-i\dcp}\\
-s_{12}c_{23}-c_{12}s_{23}s_{13}e^{i\dcp} & c_{12}c_{23}-s_{12}s_{23}s_{13}e^{i\dcp} & s_{23}c_{13}\\
s_{12}s_{23}-s_{13}c_{12}c_{23}e^{i\dcp} & -c_{12}s_{23}-s_{13}c_{23}s_{12}e^{i\dcp} & c_{23}c_{13}\\
\end{array}
\right],
\end {equation}
where $c_{ij}=\cos \theta_{ij}$ and $s_{ij}=\sin \theta_{ij}$. Thus neutrino oscillation probabilities depend on three mixing angles
$\tx$, $\ty$, and $\tz$, one CP violating phase $\dcp$, and two independent mass squared differences $\ds=m_{2}^{2}-m_{1}^{2}$ and $\dl=m_{3}^{2}-m_{1}^{2}$. 
$m_i$ is the mass of the mass eigen state $\nu_i$.  Among these parameters, $\tx$ and $\ds$ have been measured in the solar neutrino
experiments \cite{Bahcall:2004ut, Ahmad:2002jz}, $|\dl|$ and $\sin^2 2\tz$ have been measured in atmospheric neutrino experiments \cite{Fukuda:1994mc} and long base line accelerator neutrino
experiments like MINOS \cite{Kyoto2012MINOS}. The reactor neutrino experiments measured $\ty$ \cite{An:2012eh, Ahn:2012nd, Abe:2011fz}. In table \ref{bestfit}, we have noted down the present
global best-fit values of neutrino oscillation parameters. The current unknowns are the sign of $\dl$, $\dcp$ and the octant of $\tz$. Depending
on the sign of $\dl$, two scenarios are possible: a. Normal Hierarchy (NH): $m_3>>m_2>m_1$ and b. Inverted Hierarchy (IH): $m_2>m_1>>m_3$.
The long baseline accelerator neutrino experiments \nova \cite{Ayres:2004js} and T2K \cite{Itow:2001ee} are currently taking data in both neutrino and anti-neutrino
mode to observe both muon disappearances and electron appearances in the detector and are expected to measure the unknowns. 
However, their recent results \cite{Himmel:2020, Dunne:2020} show tension between the allowed regions in the $\sin^2 \tz-\dcp$ plane, with \nova excluding T2K
best-fit point at $90\%$ confidence level and T2K excluding \nova best-fit point at $3\, \sigma$. If this trend continues with the future data,
upcoming long baseline accelerator experiments like DUNE \cite{Abi:2018dnh} and T2HK \cite{Ishida:2013kba} will play vital role in resolving the tension. 

Apart from the tension in the $\sin^2 \tz-\dcp$ plane, the latest data from \nova and T2K also fail to make any distinction between the two hierarchies
\cite{Kelly:2020fkv}. Therefore, it's really important for the future experiments to distinguish between the two hierarchies. Atmospheric neutrino experiments
with multi-megaton ice/water Cherenkov detectors, like PINGU \cite{Aartsen:2014oha} and KM3NeT \cite{Adrian-Martinez:2016fdl}, can provide large statistics, however their mass hierarchy sensitivity is limited by the   
uncertainties in the flux. Careful analysis to minimise these uncertainties using dedicated detector specific simulations have been done and it has been shown that
it is possible to achieve $\simeq 3\, \sigma$ hierarchy sensitivity after 3-4 years of data taking \cite{Kouchner:2016pqa}. However the CP sensitivity is rather low unless future upgrade
is possible \cite{Razzaque:2014vba}. 

Combining multi megaton water/ice Cherenkov detector with accelerator neutrino source over a very long baseline to explore
the hierarchy and CP sensitivity has been considered in literature \cite{Fargion:2010vb, Tang:2011wn, Agarwalla:2012zu, LujanPeschard:2013ud, Brunner:2013lua, Vallee:2016xde}. In ref.~\cite{Rahaman:2017hua}, the hierarchy and CP sensitivity of a possible future experiment
with beam from Fermilab towards the ORCA detector of KM3NeT collaboration was discussed.  Ref.~\cite{Choubey:2018rnl} discussed the physics potential of an experiment
with beam from Protvino to Orca (P2O). In 2019, the letter of intent (LOI) of the P2O experiment came out \cite{Akindinov:2019flp}. The fluxes used in the LOI was
less optimistic than the ones used in ref.~\cite{Choubey:2018rnl}. Therefore it is important to explore the physics potential of the P2O with the updated beam.

Apart from measuring the standard oscillation parameter, future long baseline accelerator neutrino experiments will explore
beyond Standard Model (BSM) physics \cite{Arguelles:2019xgp}. Recently, there have been studies where BSM physics has been used to resolve the tension between the \nova and T2K. In ref.~\cite{Chatterjee:2020kkm, Denton:2020uda} non-standard neutral current interaction during propagation has been used to resolve this tension, while in ref.~\cite{Rahaman:2021leu} same has been done with CPT violating Lorentz invariance violation. Ref.~\cite{Miranda:2019ynh} tried to probe non-unitary mixing with the recent \nova and T2K data. They found out that the tension between the two experiments can be reduced with non-unitarity and that both the experiments prefer non-unitary mixing. They also showed that the signature of non-unitarity in these two experiments have grown stronger over the years. However, a recent paper \cite{Forero:2021azc} showed that the short baseline experiments NOMAD and NuTeV prefer unitary mixing and put strong constraints on non-unitary parameters. This means that the short baseline experiments data have a tension with the long baseline data from the \nova and T2K. Thus it is important to explore the effects of non-unitary mixing with different baselines and energy.

Earlier, effects of non-unitary parameters on the determination of unknown neutrino oscillation parameters in long baseline experiments like NO$\nu$A, T2K, DUNE, T2HK and TNT2K have been discussed in ref.~\cite{Dutta:2016czj, Dutta:2016eks, Dutta:2016vcc, Ge:2016xya, Ge:2017qqv}. Ref.~\cite{Dutta:2019hmb} explored the role of neutral current measurements in measuring the bounds on non-unitary parameters. 
In this paper, we have studied the effect of non-unitary mixing on the determination of hierarchy and CP violation in P2O experiment. In section~\ref{P2O}, we have discussed the detail of the P2O experiment. Neutrino oscillation probabilities with unitary and non-unitary mixing have been discussed in section~\ref{prob}. The physics sensitivities to determine hierarchy both with unitary and non-uniray oscillation have been talked about in section~\ref{hierarchy}. Section~\ref{CPV} talks about the physics potential to establish CP violation in the P2O experiment with unitary and non-unitary mixing schemes. The future constraints on non-unitary parameters from P2O experiment have been discussed in section \ref{const}. The conclusion has been drawn in section~\ref{conclusion}. 

\section{Experimental details}
\label{P2O}
ORCA (Oscillation Research with Cosmics in the Abyss) is the low energy component of the KM3NeT Consortium \cite{Adrian-Martinez:2016fdl}, housing two detectors in the Mediterranean sea. It is located about $40$ km off the coast of Toulon, France, at a depth of $\sim 2450$ m. Upon completion ORCA will house 115 strings with an effective volume of $\sim 5$ MT. The water Cherenkov detector is optimized for the study of atmospheric neutrino oscillations in the energy range of 3 to 100 GeV with the primary goal to
determine the neutrino mass ordering. 
%The majority of neutrino events observed by ORCA will be from electron and muon neutrino and antineutrino charge-current (CC) interactions, while tau neutrinos and neutral current (NC) interactions constitute minor backgrounds. 
A $3\,\sigma$ sensitivity to mass hierarchy is expected after 3 years of full-ORCA data taking \cite{Akindinov:2019flp}. ORCA will provide a complimentary measurement of $\Delta_{32}$ and $\theta_{23}$, measure the $\nu_\tau$ flux normalisation, and probe a variety of new physics scenarios namely, sterile neutrinos, non-standard interactions, neutrino decay and so on. Currently, 6 ORCA lines are live and taking data since an year.

The Protvino accelerator complex, situated at $100$ km south of Moscow, sits a U-70 accelerator. The baseline between Protvino and ORCA is roughly $2595$ km. 
%The present beam power is $15$ kW. 
A beam power of $90$ kW, corresponding to $0.8\times 10^{20}$ protons on target (POT) per year \cite{Akindinov:2019flp} is expected. However, there is a proposal to upgrade the beam power to $450$ kW \cite{Akindinov:2019flp}. 

In ref.~\cite{Choubey:2018rnl}, the analyses have been done assuming an optimistic beam power of $450$ kW. In this work, we have taken a conservative approach followed in the P2O LOI~\cite{Akindinov:2019flp}. To do so, we have first matched the expected event numbers with the LOI \cite{Akindinov:2019flp}. To calculate event numbers, the GLoBES \cite{Huber:2004ka, Huber:2007ji} simulation package has been used. For energy smearing, GLoBES incorporates a Gaussian resolution function:
\begin{equation}
R^c (E,E^\prime)=\frac{1}{\sqrt{2\pi}}e^{-\frac{(E-E^\prime)^2}{2\sigma^2(E)}},
\end{equation}
where $E$ is the true neutrino energy and $E^\prime$ is the reconstructed neutrino energy. The energy resolution is parameterised by:
\begin{equation}
\sigma(E)=\alpha E+\beta \sqrt{E}+\gamma.
\end{equation}
Values of $\alpha=0$, $\beta=0$, and $\gamma=0.4$ are assumed in this work.

Depending on the Cherenkov signatures of the outgoing lepton from the $\nu_{e}$ and $\nu_{\mu}$ CC and NC interactions, two distinct event topologies are observed at the detector: track-like and shower-like events. $\nu_{\mu}$ CC and those $\nu_{\tau}$ CC interactions with muonic $\tau$ decays mostly account for track-like topology. The shower-like topology has events from $\nu_{e}$ CC, $\nu_{\tau}$ CC interactions with non-muonic $\tau$ decays  and NC interactions of all flavours.  

\begin{figure}[H]
  \centering
  \includegraphics[width=0.49\textwidth]{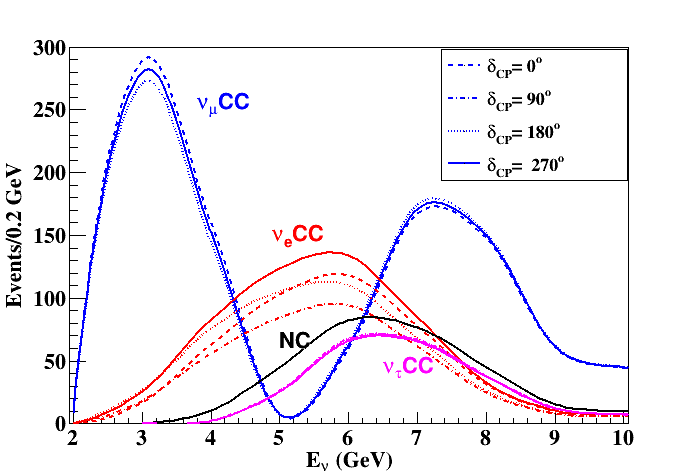}
    \includegraphics[width=0.49\textwidth]{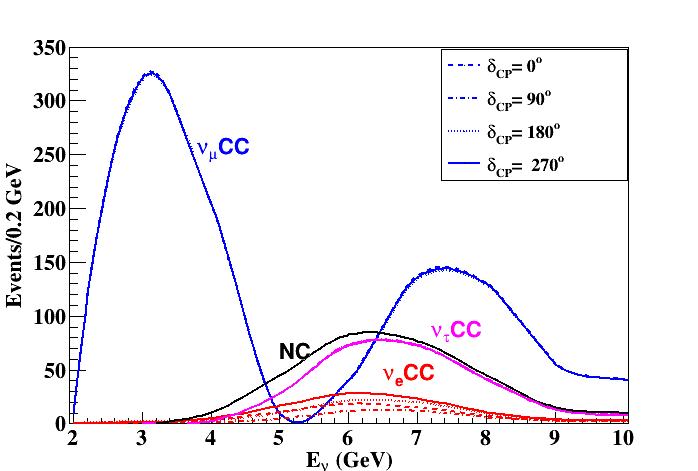}
    \caption{\label{evt_plots} Energy  distribution  of  the  expected  number  of  beam neutrino  events  that  would  be detected by ORCA after 3 years of running  for 4 different values of the CP phase for the case of normal (left panel) and inverted (right panel)  neutrino  mass  ordering. $\theta_{23}=45^{o}$ is  assumed. The  x-axis  shows the  true neutrino energy.}
    \label{events}
\end{figure}

For the disappearance channels, the signal comes from $\nu_\mu$ CC events, and the backgrounds come from NC and tau events. For appearance channels, the signal comes from $\nu_e$ CC events and the backgrounds come from $\nu_\tau$ CC events, NC events and also the mis-identification. In our analysis, we have modified GLoBES to match the expected event rates given in ref.~\cite{Akindinov:2019flp}. The energy distributions of the expected neutrino events across various channels after 3 years of running in the neutrino mode (corresponding to $2.4\times 10^{20}$ POT) are shown in Figure~\ref{events}. 

\section{Oscillation probabilities with unitary and non-unitary mixing}
\label{prob}
The standard unitary neutrino oscillation probability with matter effect for uniform matter density can be written as \cite{Cervera:2000kp}
\begin{eqnarray}
  \pme &\simeq& \sin^2 2 \ty \sin^2 \tz\frac{\sin^2\dhat(1-\ahat)}{(1-\ahat)^2}\nonumber\\
  &+& \alpha \cos \ty \sin2\tx \sin 2\ty \sin 2\tz \cos(\dhat+\dcp)
  %\nonumber\\ &&
 \frac{\sin\dhat \ahat}{\ahat}
  \frac{\sin \dhat(1-\ahat)}{1-\ahat},
  \label{pme}
   \end{eqnarray}
where $\alpha=\frac{\ds}{\dl}$, $\dhat=\frac{\dl L}{4E}$ and $\ahat=\frac{A}{\dl}$. $A$ is the Wolfenstein matter term \cite{msw1}, given by $A=2\sqrt{2}G_FN_eE$, where $E$ is the neutrino beam energy and $L$ is the length of the baseline. Anti-neutrino oscillation probability $\pmebar$ can be obtained by changing the sign of $A$ and $\dcp$ in eq.~\ref{pme}. The oscillation probability mainly depends on hierarchy (sign of $\dl$), octant of $\tz$ and $\dcp$, and precision in the value of $\ty$. This dependency leads to eight fold degeneracy \cite{Fogli:1996pv, BurguetCastell:2001ez, Minakata:2001qm, Mena:2004sa, Prakash:2012az, Meloni:2008bd, Agarwalla:2013ju, Nath:2015kjg, Bora:2016tmb, Barger:2001yr, Kajita:2006bt}. Since the Reactor neutrino experiments \cite{An:2012eh, Ahn:2012nd, Abe:2011fz} have measured $\ty$ precisely and thus broke the eight fold degeneracy into $1+3+3+1$ pattern \cite{Bharti:2018eyj, Rahaman:2022rfp}. 

\begin{figure}[H]
  \centering
  \includegraphics[width=0.8\textwidth]{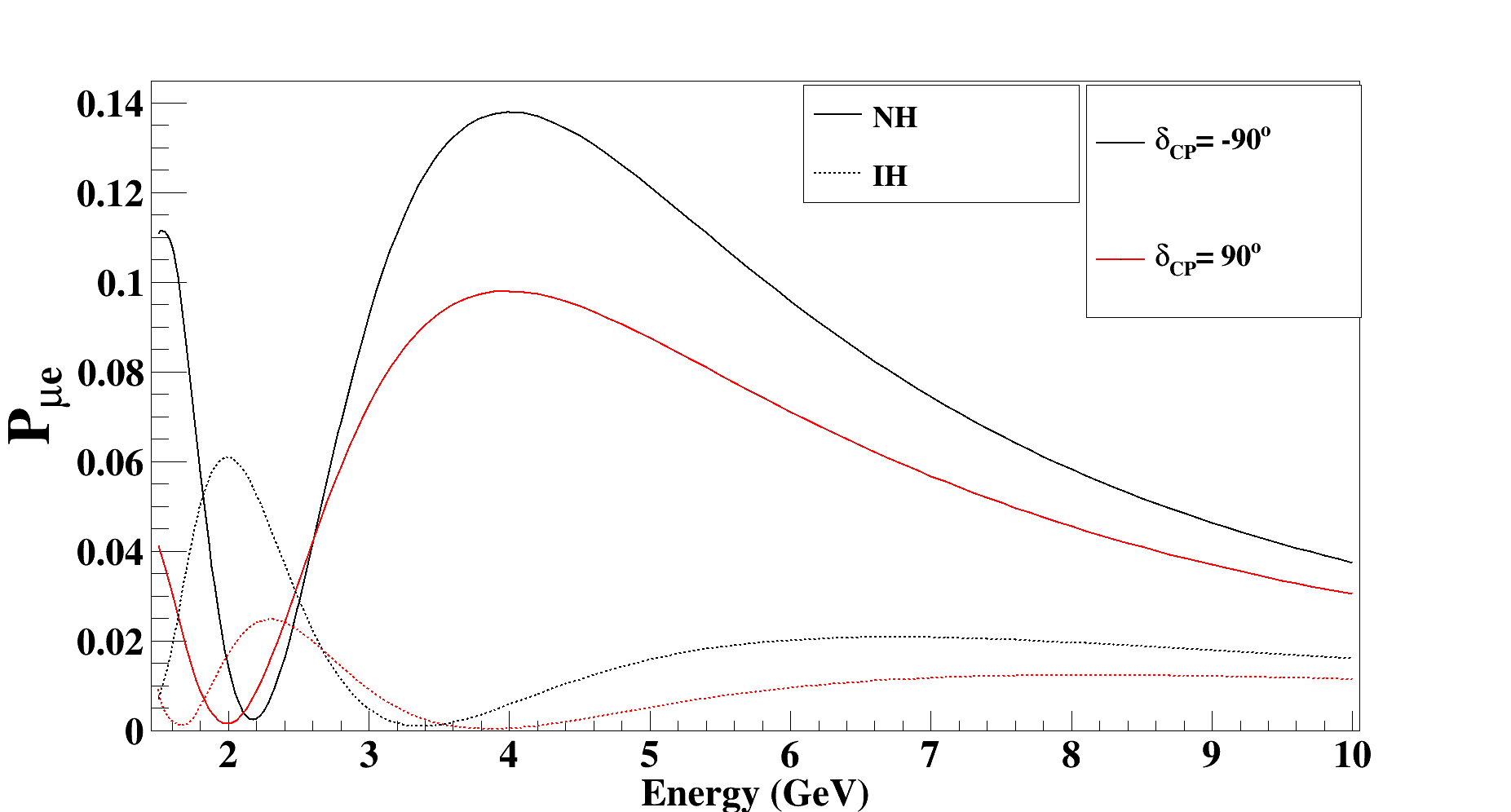}
  \includegraphics[width=0.8\textwidth]{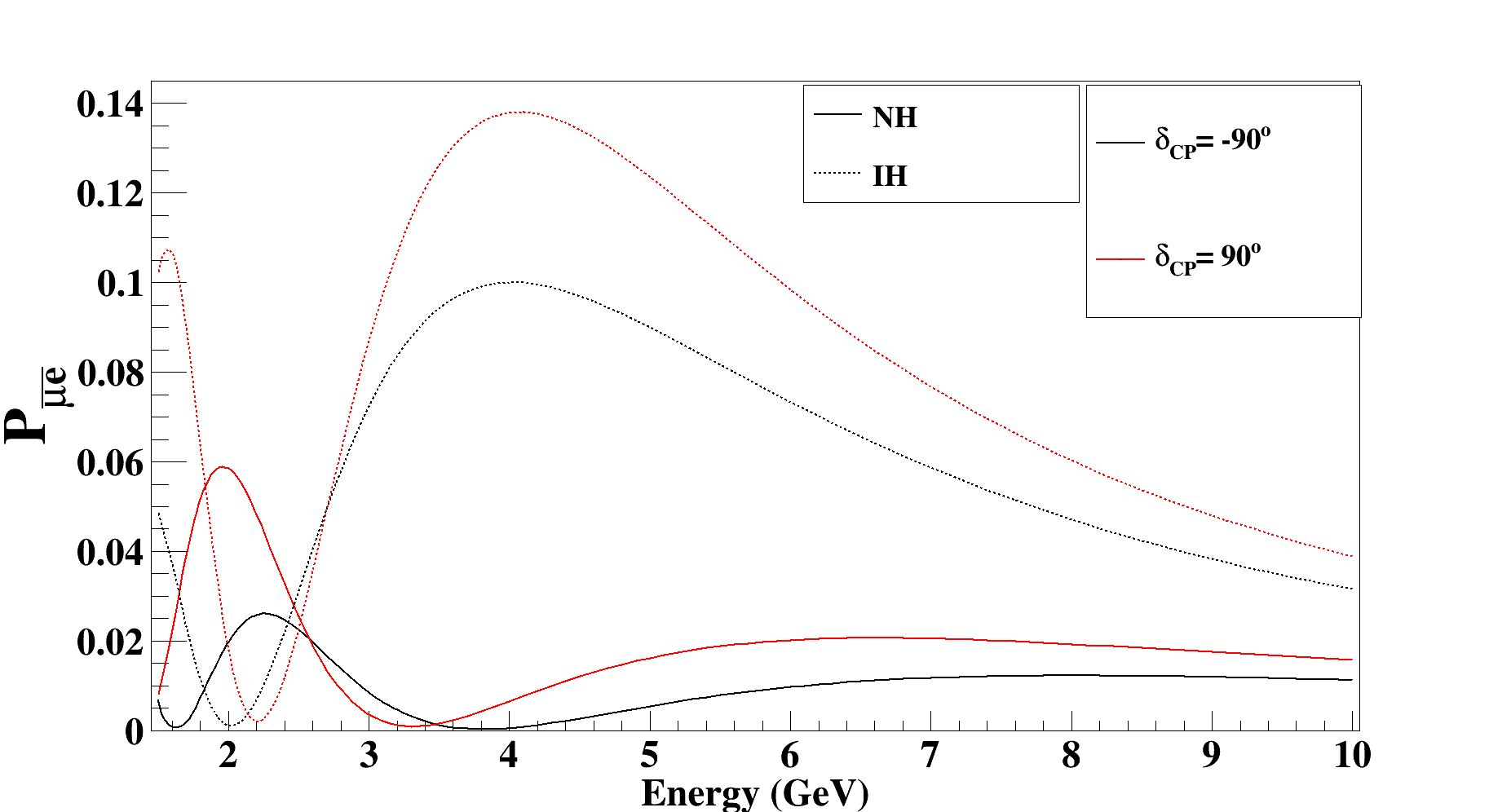}
   \caption{ (Anti-) Neutrino appearance probability as a function of energy in the upper (lower) panel assuming NH and IH with standard unitary oscillation parameters given in Table~\ref{bestfit} for $\delta_{cp}=-90^{\circ}$ and $\delta_{cp}=+90^{\circ}$}
   \label{prob-uni}
\end{figure}

The hierarchy-$\dcp$ degeneracy is important for experiments like \nova and T2K, making it impossible to determine the hierarchy when NH (IH) and $0<\dcp<180^\circ$ ($-180^\circ<\dcp<0$) is the true hierarchy-$\dcp$ combination chosen by nature. We call $0<\dcp<180^\circ$ upper half plane (UHP), and $-180^\circ<\dcp<0$ lower half plane (LHP) of $\dcp$. In eq.~\ref{pme}, NH (IH) enhances (suppresses) $\pme$, while $\dcp$ in UHP (LHP) suppresses (enhances) $\pme$. Thus the value of probabilities for NH and $\dcp$ in UHP, and IH and $\dcp$ in LHP are close to each other, making these two combinations as the unfavourable hierarchy-$\dcp$ combinations. On the other hand, for the combinations NH-$\dcp$ in LHP, and IH-$\dcp$ in UHP, the values of $\pme$ are well separated. These two are the favourable hierarchy-$\dcp$ combinations. Hierarch-$\dcp$ degeneracy exists for $\pmebar$ as well. However, this degeneracy can be broken in the case of bi-magic baseline 2540 km, as shown in ref.~\cite{Raut:2009jj}. The P2O baseline $2595$ km is very close to the bi-magic baseline, and hence the special effects enhancing the hierarchy sensitivity of bi-magic baseline are also present in the case of P2O as well. In fig.~\ref{prob-uni}, we have shown the oscillation probabilities $\pme$ and $\pmebar$ as a function of energy for P2O and for different hierarchy and $\dcp$ values as mentioned in the plot. The other oscillation parameters have been fixed at values mentioned in table~\ref{bestfit}. We have only plotted probabilities pertaining to $\dcp$ values $-90^\circ$ and $90^\circ$. All other $\dcp$ values fall in between. It can be easily noticed that the hierarchy-$\dcp$ degeneracy gets broken and hierarchy can be determined even for the unfavourable hierarchy-$\dcp$ combinations. However, for $\pme$, the separation between NH-$\dcp=+90^\circ$, and IH-$\dcp=-90^\circ$ is least compared to other hierarchy-$\dcp$ combinations. But for $\pmebar$, the separation between these two hierarchy-$\dcp$ combinations is maximum. Thus, in case of P2O addition have anti-neutrino run will have better hierarchy sensitivity than only neutrino run for NH-$\dcp$ in UHP, and IH-$\dcp$ in LHP.

\begin{figure}[H]
  \centering
  \includegraphics[width=0.8\textwidth]{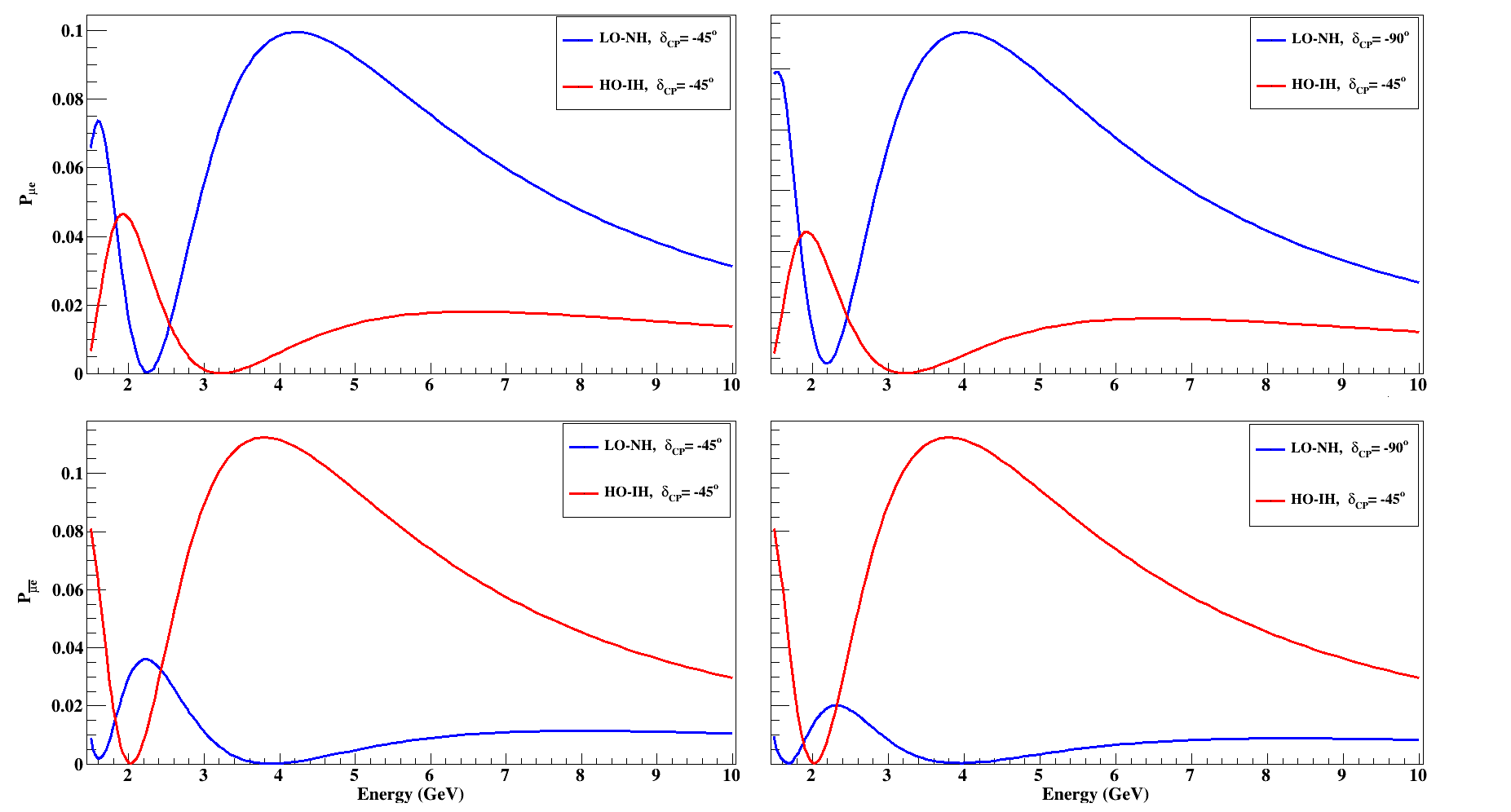}
   \caption{$\nu_\mu \to \nu_e$ probability plots for LO-NH, and HO-IH and for different $\dcp$ values. The upper (lower) panel is for neutrino (anti-neutrino) probability. We considered $\sin^2\tz=0.45$ ($0.57$) for LO (HO).}
   \label{prob-uni-oct}
\end{figure}

Another degeneracy affecting the hierarchy sensitivity of the long baseline accelerator neutrino experiments is the octant-hierarchy sensitivity \cite{Prakash:2013dua, Rahaman:2022rfp}. As explained in ref.~\cite{Prakash:2013dua, Rahaman:2022rfp}, $\pme$ gets a positive (negative) boost due to both hierarchy and octant when the hierarchy is NH (IH) and the octant of $\tz$ is HO (LO). Therefore, HO-NH and LO-IH are well separated from each other and they are the favourable octant-hierarchy combinations. When the hierarchy is NH (IH), and $\tz$ is in LO (HO), $\pme$ gets a positive (negative) boost due to hierarchy and negative (positive) boost due to octant. Therefore LO-NH and HO-IH are two unfavourable octant-hierarchy combinations in the long baseline experiments like \nova and T2K. This octant-hierarchy degeneracy can be broken down by anti-neutrino run.  However, as we can see from fig.~\ref{prob-uni-oct} that the in case of P2O, octant-hierarchy degeneracy is not present even for the neutrino run. This is also because of the magic-baseline property of P2O. But even then, anti-neutrino run provides better separation between the unfavourable octant-hierarchy combinations. To generate the probability plots, we considered $\sin^2\tz=0.45$ ($0.57$) for LO (HO). 

In the case of non-unitary mixing, the mixing matrix is defined as \cite{Escrihuela:2015wra} 
\begin{equation}
N=N_{NP}U_{3\times 3}= \left[ {\begin{array}{ccc}
   \alpha_{00} & 0 & 0 \\
   \alpha_{10} & \alpha_{11} & 0 \\
   \alpha_{20} & \alpha_{21} & \alpha_{22}
  \end{array} } \right] U_{\rm PMNS}.
\end{equation}
The mixing between flavour states and mass eigenstates can be written as
\begin{equation}
    |\nu_{\beta}>=\sum_{i=1}^{3}N_{\beta i}|\nu_i>,
\end{equation}
where $\beta$ denotes the flavour states and $i$ denotes the mass eigenstates. The evolution of neutrino mass eigenstates during the propagation, can be written as
\begin{equation}
    i\frac{d}{dt}|\nu_i>=H_{\rm vac}|\nu_i>,
\end{equation}
where $H_{\rm vac}$ is the Hamiltonian in vacuum and it is defined as
\begin{equation}
    H_{\rm vac}=\frac{1}{2E}\left[ {\begin{array}{ccc}
   0 & 0 & 0 \\
   0 & \ds & 0 \\
   0 & 0 & \dl
  \end{array} } \right]
\end{equation}
The non-unitary neutrino oscillation probability in vacuum can be written as
\begin{eqnarray}
    \pme^{\rm NU}({\rm vac})&=& \sum_{i,j}^{3}N_{\mu i}^{*}N_{ei}N_{\mu j}N_{ej}^{*}-4\sum_{j>i}^{3}Re\left[N_{\mu j}N_{ej}N_{\mu i}N_{ei}^{*}\right]\sin^2 \left(\frac{\Delta_{ji}L}{4E}\right)\nonumber \\
    &&+2\sum_{j>i}^{3}Im\left[N_{\mu j}^{*}N_{ej}N_{\mu i}N_{ei}\right]\sin \left(\frac{\Delta_{ji}L}{2E}\right).
    \label{pme-nu}
\end{eqnarray}
If written explicitly neglecting the cubic products of $\alpha_{10}$, $\sin \ty$, and $\ds$, the oscillation probability takes the form \cite{Escrihuela:2015wra}
\begin{equation}
    \pme^{\rm NU}({\rm vac})= (\alpha_{00}\alpha_{11})^2\pme^{SO}+\alpha_{00}^2\alpha_{11}|\alpha_{10}|\pme^I+\alpha_{00}^2|\alpha_{10}|^2.
\end{equation}
$\pme^{SO}$ is the standard three flavour unitary neutrino oscillation probability in vacuum and can be written as
\begin{eqnarray}
    \pme^{SO}&=& 4\cos^2 \tx \cos^2 \tz \sin^2 \tx\sin^2\left(\frac{\ds L}{4E}\right)\nonumber \\
    &&+ 4\cos^2 \ty \sin^2 \ty \sin^2 \tz\sin^2\left(\frac{\dl L}{4E}\right)\nonumber \\
    &&+ \sin (2\tx)\sin \ty \sin (2\tz)\sin\left(\frac{\ds L}{2E}\right)\sin\left(\frac{\dl L}{4E}\right)\cos\left(\frac{\dl L}{4E}-I_{123}\right),
   \end{eqnarray}
and
\begin{eqnarray}
    \pme^I&=&-2\left[\sin (2\ty)\sin\tz\sin \left(\frac{\dl L}{4E}\right)\sin\left(\frac{\dl L}{4E}+I_{NP}-I_{012}\right)\right]\nonumber \\
    &&-\cos \ty \cos \tz \sin (2\tx)\sin\left(\frac{\ds L}{2E}\right)\sin (I_{NP}),
\end{eqnarray}
where $I_{012}=-dcp=\phi_{10}-\phi_{20}+\phi_{21}$, and $I_{NP}=\phi_{10}-{\rm Arg}(\alpha_{10})$. $\phi_{ij}$s are the phases associated with $\alpha_{ij}=|\alpha_{ij}|e^{i \phi_{ij}}$. It should be noted that non-unitary parameters $\alpha_{00}$, $\alpha_{11}$, and $\alpha_{10}$ have most significant effects on $\pme^{\rm NU}({\rm vac})$.

While, propagating through matters, the neutrinos undergo forward scattering and the neutrino oscillation probability gets modified due to interaction potential between neutrino and matter. In case of non-unitary mixing, the CC and NC interaction Lagangian is given as
\begin{equation}
    \mathcal{L}=V_{\rm CC}\sum_{i,j}N_{ei}^{*}N_{ej}\bar{\nu}_i\gamma^0\nu_j+V_{\rm NC}\sum_{\alpha,i,j}N_{\alpha i}^{*}N_{\alpha j}\bar{\nu}_i\gamma^0\nu_j,
\end{equation}
where $V_{\rm CC}=\sqrt{2}G_FN_e$, and $V_{\rm NC}=-G_F N_n/\sqrt{2}$ are the potentials for CC and NC interactions respectively. Therefore, the effective Hamiltonian becomes
\begin{equation}
    H_{\rm matter}^{\rm NU}= \frac{1}{2E}\left[ {\begin{array}{ccc}
   0 & 0 & 0 \\
   0 & \ds & 0 \\
   0 & 0 & \dl
  \end{array} } \right]+N^\dagger \frac{1}{2E}\left[ {\begin{array}{ccc}
   V_{\rm CC}+V_{\rm NC} & 0 & 0 \\
   0 & V_{\rm NC} & 0 \\
   0 & 0 & V_{\rm NC}
  \end{array} } \right]N
\end{equation}
The non-unitary neutrino oscillation probability after neutrinos travel through a distance $L$ is given as
\begin{equation}
    P_{\alpha \beta} (E,L)= |<\nu_\beta|\nu_\alpha (L)>|^2=\left|\left(Ne^{-iH_{\rm matter}^{\rm NU}L}N^\dagger\right)_{\beta \alpha}\right|^2
\end{equation}
A detailed description of non-unitary neutrino oscillation probability in presence of matter effect has been discussed in ref.~\cite{Miranda:2019ynh, delAguila:2002sx, Bekman:2002zk, Goswami:2008mi}.  

\begin{figure}[H]
  \centering
  \includegraphics[width=\textwidth]{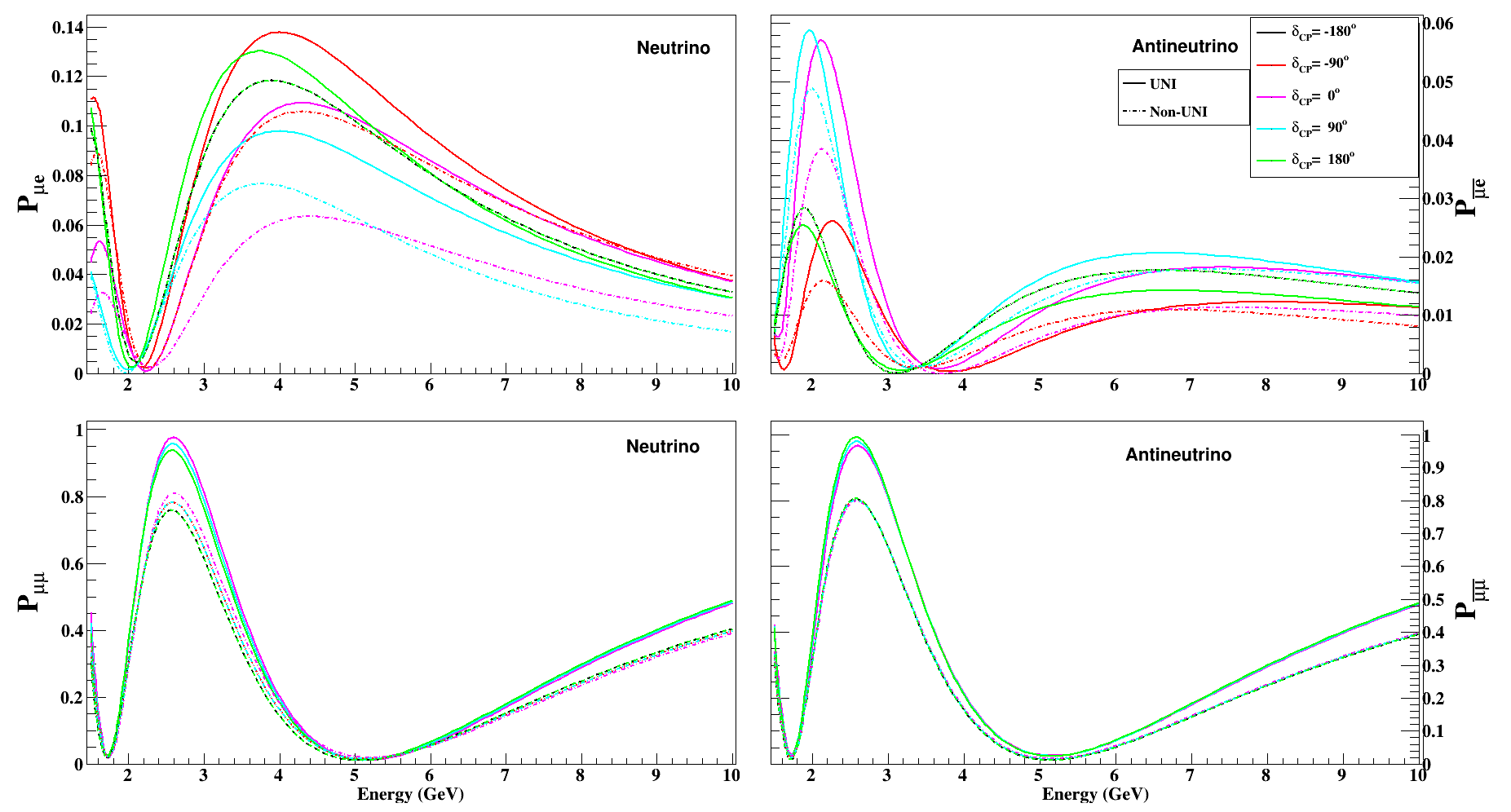}
   \caption{ Neutrino and Anti-neutrino appearance (top row) and disappearance (bottom row) probability plots assuming NH with standard oscillation parameters and with non-unitary parameters as given in Table~\ref{values} for $\phi_{10}=0^{o}$ for different true values of $\delta_{cp}$}
   \label{prob1}
\end{figure}

\begin{figure}[H]
\centering
  \includegraphics[width=1\textwidth]{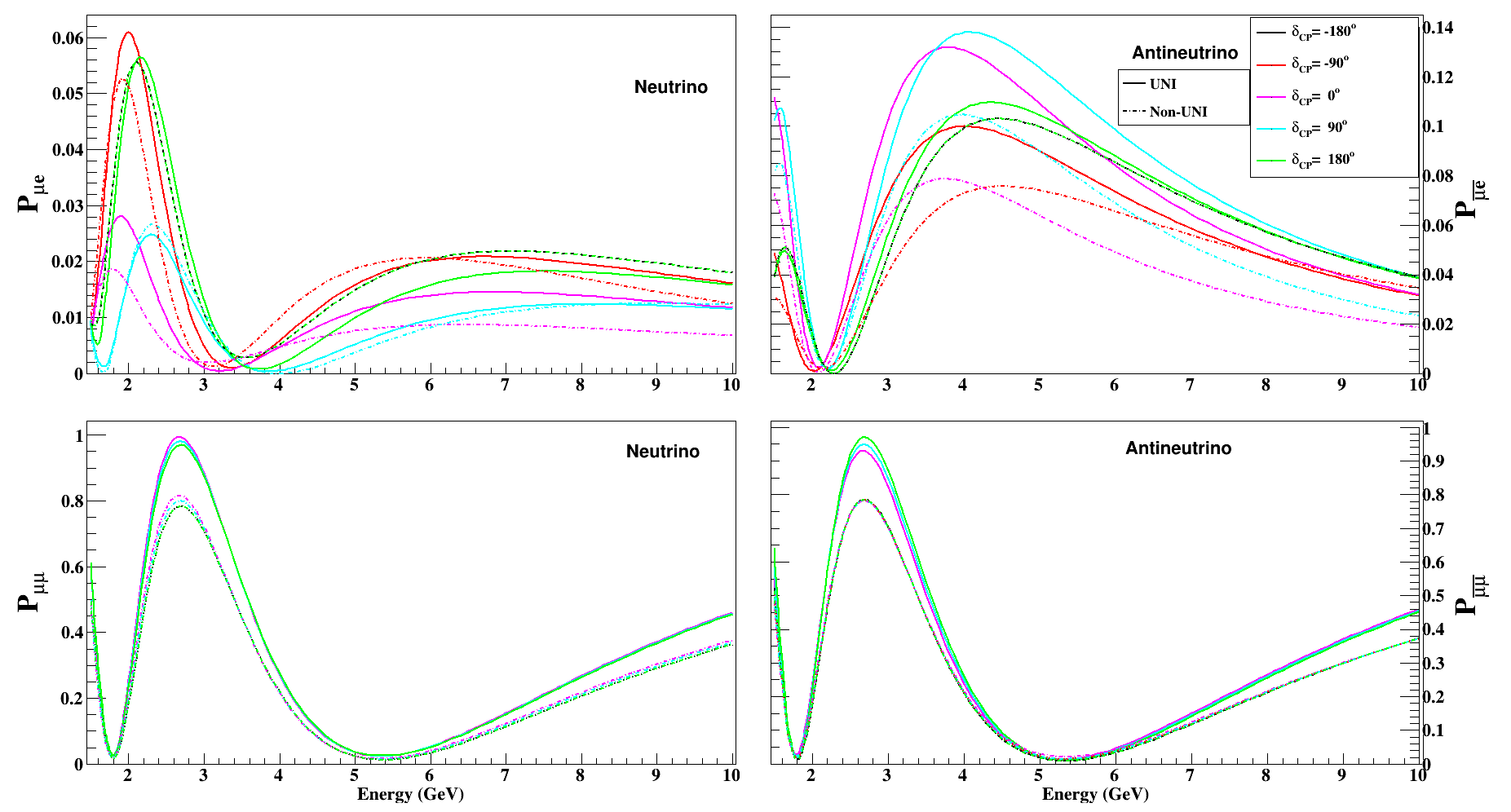}
\caption{Same as fig.~\ref{prob1}, except for IH.}
\label{prob2}
   \end{figure}

\begin{figure}[H]
  \centering
  \includegraphics[width=1.\textwidth]{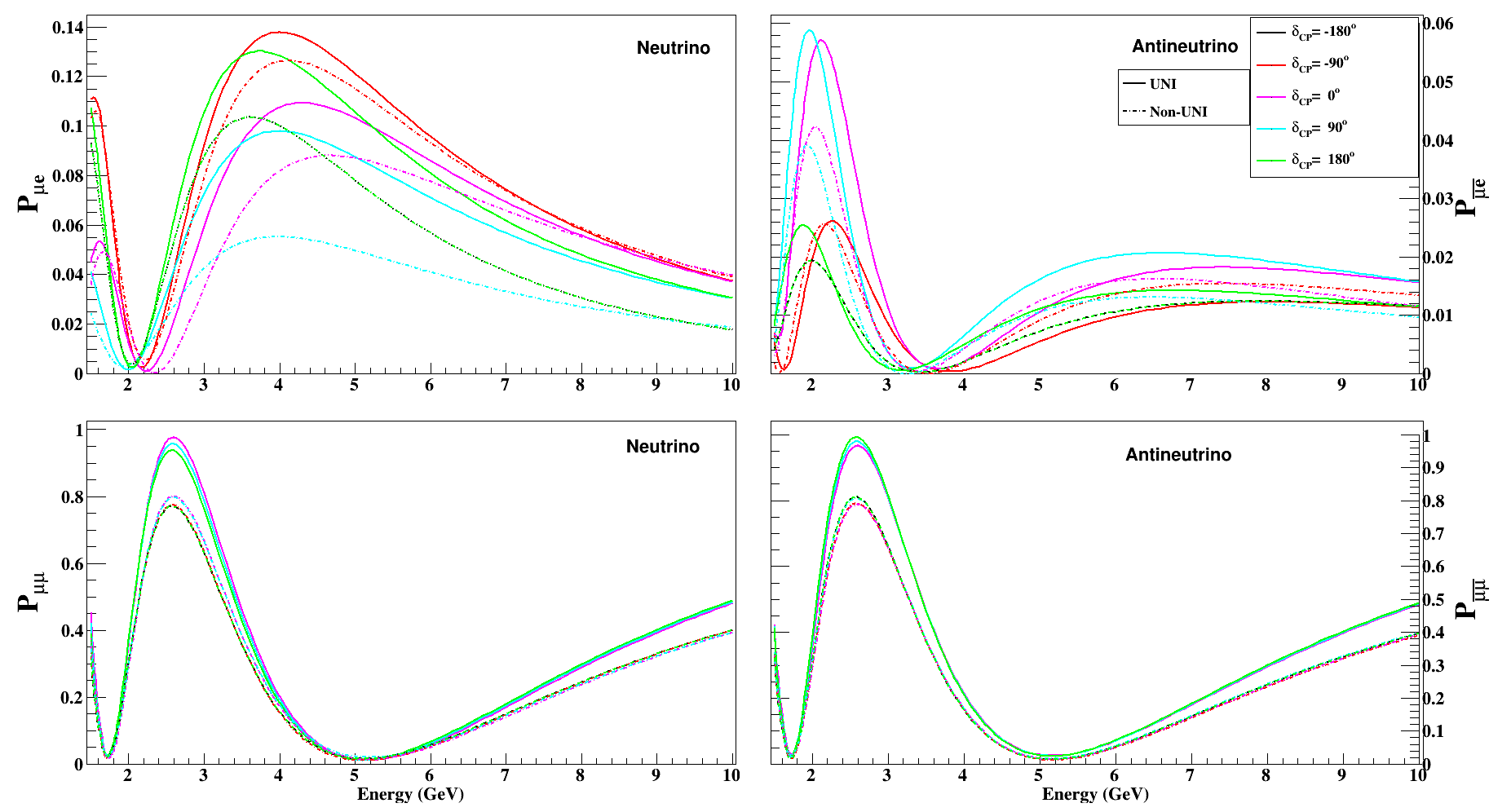}
  \caption{Neutrino and Anti neutrino appearance (top row) and disappearance (bottom row) probability plots assuming NH with standard oscillation parameters and with non-unitary parameters as given in Table~\ref{values} for $\phi_{10}=90^{o}$ for different true values of $\delta_{cp}$}
    \label{prob3}
\end{figure}

\begin{figure}[H]
     \centering
     \includegraphics[width=1.\textwidth]{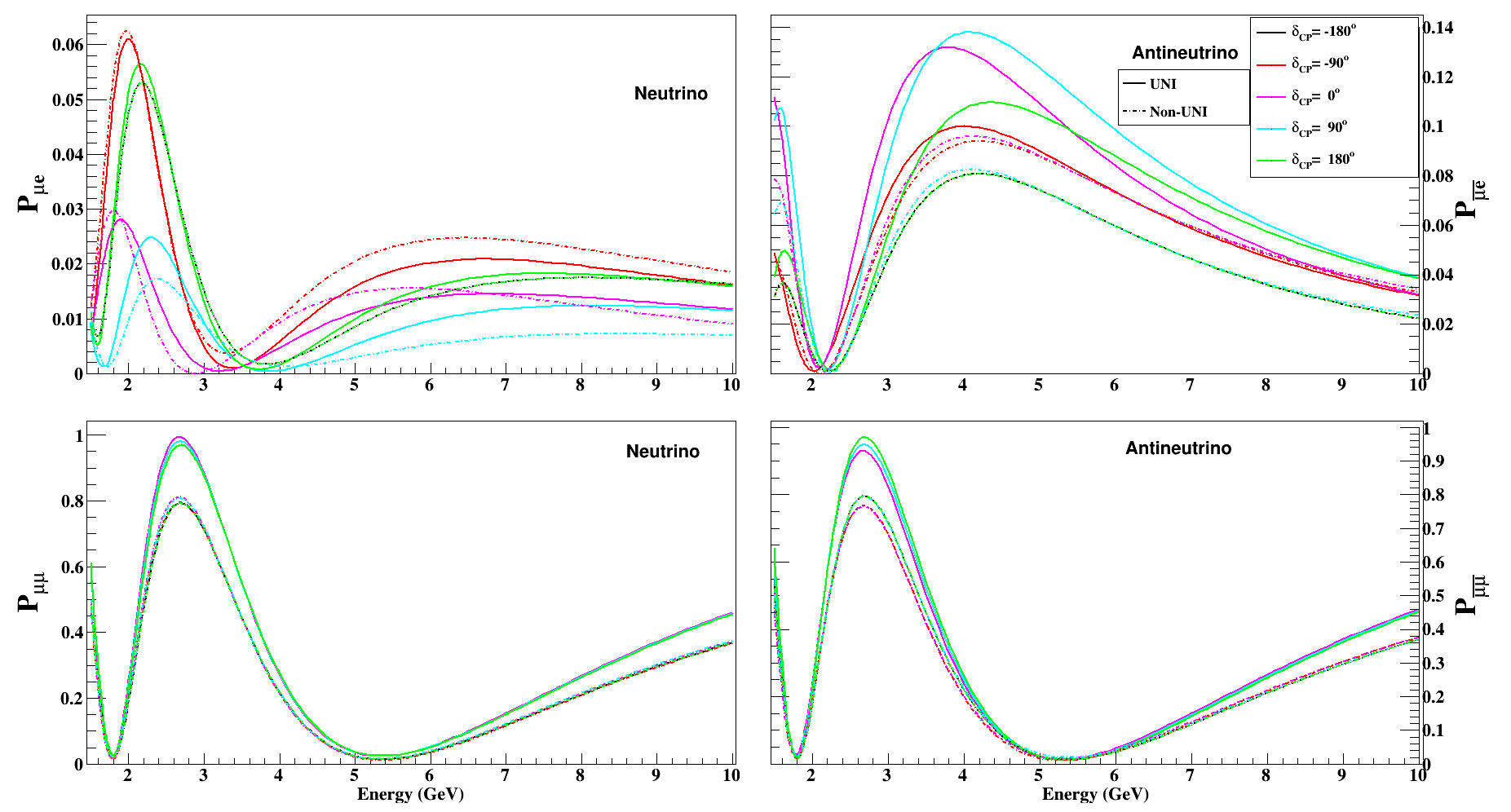}
     \caption{Same as fig.~\ref{prob3}, except for IH.}
     \label{prob4}
 \end{figure}

\begin{figure}[H]
  \centering
 \includegraphics[width=1.\textwidth]{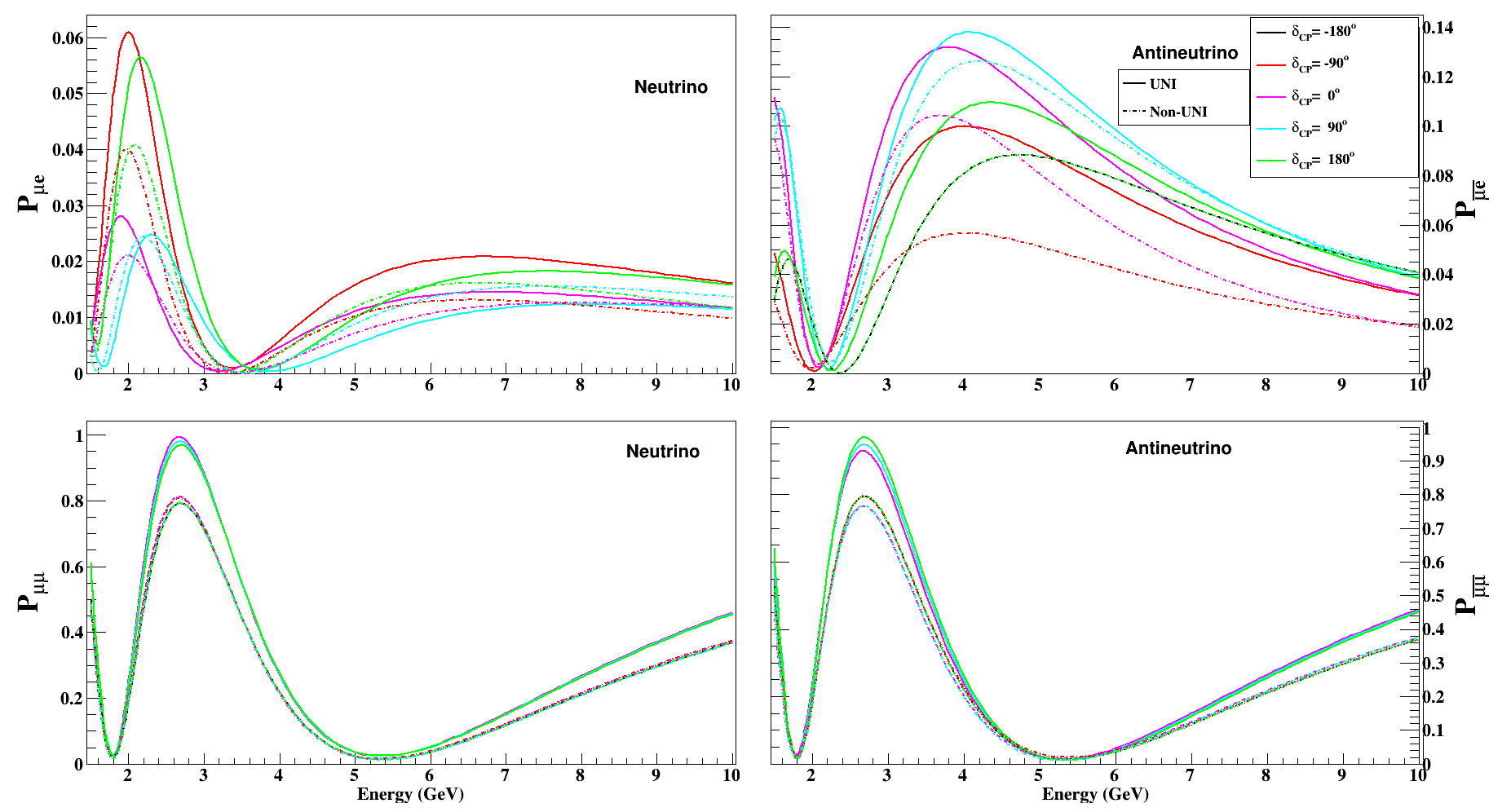}
  \caption{Neutrino and Anti neutrino appearance (top row) and disappearance (bottom row) probability plots assuming NH with standard oscillation parameters and with non-unitary parameters as given in Table~\ref{values} for $\phi_{10}=-90^{o}$ for different true values of $\delta_{cp}$}
    \label{prob5}
\end{figure}

\begin{figure}[H]
     \centering
     \includegraphics[width=1.\textwidth]{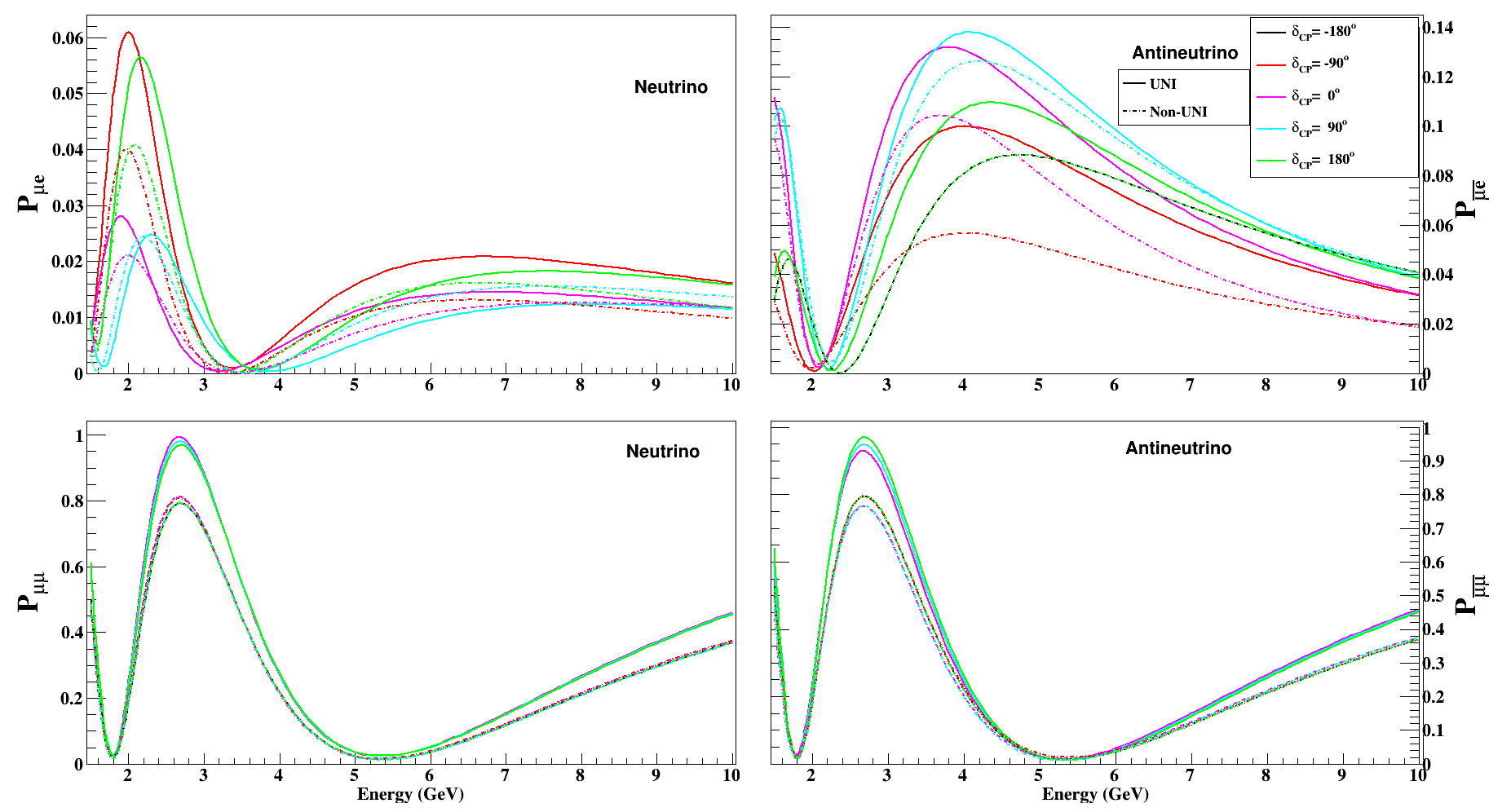}
     \caption{Same as fig.~\ref{prob5}, except for IH.}
     \label{prob6}
 \end{figure}

Probability plots for P2O with the standard and with non-unitary oscillations have been plotted for [$\phi_{10}$ =0, +90 and -90 degree] for different values of $\delta_{cP}$ under Normal hierarchy and Inverted hierarchy for neutrinos and anti-neutrinos separately. Oscillation parameters used for probability plots are given in table~\ref{values}. Please note that apart from $\alpha_{00}$, $\alpha_{10}$, and $\alpha_{11}$, all other non-unitary parameters have been fixed at their unitary values. This is because $\alpha_{00}$, $\alpha_{10}$, and $\alpha_{11}$ have maximum effects on the oscillation probability, and in this paper, we will concentrate on the effects of these three parameters on the physics potential of P2O. We have used the software GLoBES \cite{Huber:2004ka, Huber:2007ji} to calculate both unitary and non-unitary oscillation probabilities. For non-unitary oscillation probabilities, GLoBES has been modified aptly to include non-unitary mixing. Figure~\ref{prob1},~\ref{prob2} and ~\ref{prob3} shows the neutrino and anti neutrino probability plots corresponding to $\phi_{10}$ =0, +90 and -90 degree respectively.
\begin{table}[H]
\begin{center}
\begin {tabular}{|c| c |c|}

\hline
Standard osc. parameters & NH & IH \\ \hline
$\Delta_{32}$ (eV$^{2}$) & 2.52$e^{-3}$ &  2.512$e^{-3}$\\
$\Delta_{21}$ (eV$^2$) & 7.6$e^{-5}$ & 7.6$e^{-5}$  \\
$ \sin^2\theta_{23}$ & 0.582 & 0.582 \\
$ \sin^2\theta_{13}$ & 0.02240 & 0.02263 \\
$ \delta_{CP}$ (degree) & 0,90,180,360 & 0,90,180,360  \\
\hline
Non unitary parameters & (both NH and IH)& \\ \hline
$\alpha_{00}$ &0.93 & \\
$\alpha_{10}$  & $3.6e^{-2}$ & \\
$\alpha_{11}$ & 0.95 & \\
$\alpha_{20}$ & 0 &  \\
$\alpha_{21}$ & 0 &  \\
$\alpha_{22}$ & 1.0 &  \\
$ \phi_{10}$ & 0,+90,-90 &  \\
$ \phi_{20}$ & 0 &    \\
$ \phi_{21}$ & 0 &    \\
\hline
\hline
\hline 
\end {tabular}
\caption{The neutrino and anti-neutrino oscillation parameters and their range used for probability plots} 
\label{values}
\end{center}

\end{table}
From figures~\ref{prob1} and \ref{prob2}, it is evident that when $\phi_{10}$=0, P2O has good discrimination potential between unitary and non-unitary mixing at the peak energy in both appearance and disappearance channels, for both neutrino and anti-neutrino mode and for all the values of $\dcp$ and both the hierarchies. When $\phi_{10}=90^\circ$ ($-90^\circ$), P2O cannot discriminate between unitary and non-unitary mixing for $\dcp=-90^\circ$ ($90^\circ$) in the appearance channel for both neutrino and anti-neutrino modes, when NH is the true hierarchy. However, for all other $\dcp$ values, the discrimination potential between the two mixing schemes is quite good for both the channels, modes and hierarchies. These features have been illustrated in figures \ref{prob3}-\ref{prob6}. 

.

\section {Mass hierarchy sensitivity}
\label{hierarchy}
In this section, we will discuss the mass hierarchy sensitivity of the P2O experiment for different neutrino and anti-neutrino run times. From now on, $x$ years of neutrino run and $y$ years of anti-neutrino run will be defined as $x+y$ years. 
\subsection{Mass hierarchy sensitivity with standard unitary oscillation}
\label{hierarchy-uni}
 First we have considered that NH is our true hierarchy and IH is test hierarchy. For the true data set, values of $\delta_{cp}$ has been varied in the range $[-180^\circ:180^\circ]$. Other standard unitary oscillation parameter values have been fixed at the values given in ref.~\cite{nufit, Esteban:2020cvm}. We have assumed unitary mixing for both true and test event rates. In the test data set $\dl$, and $\sin^2\theta_{23}$ have been varied in their possible $3\, \sigma$ range given in ref.~\cite{nufit, Esteban:2020cvm}, $\dcp$ has been varied in its complete range $[-180^\circ:180^\circ]$. The $\chi^2$ between true and test event rates have been calculated using GLoBES \cite{Huber:2004ka, Huber:2007ji}. Automatic bin based energy smearing for generated test events has been implemented in the same way as described in the GLoBES manual \cite{Huber:2004ka, Huber:2007ji}.
 
 We used  $5\%$ normalisation and $3\%$ energy calibration systematics uncertainty for $e$ and $\mu$ like events \cite{Akindinov:2019flp}.
 Implementing systematics uncertainty has been discussed in details in GLoBES manual \cite{Huber:2004ka, Huber:2007ji}.

\begin{figure}[H]
  \centering
 \includegraphics[width=0.9\textwidth]{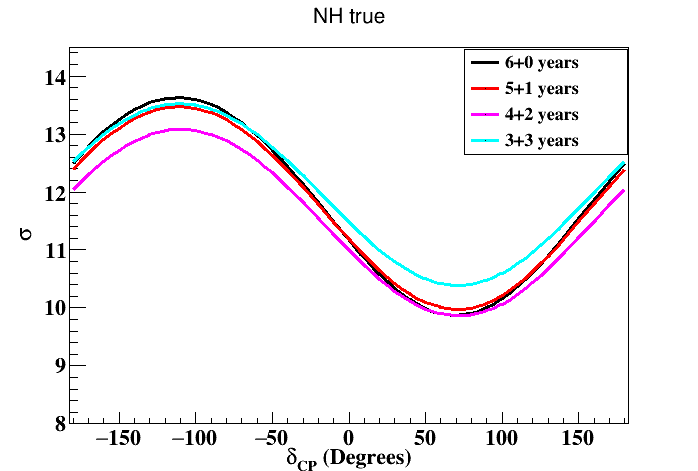}
\includegraphics[width=0.9\textwidth]{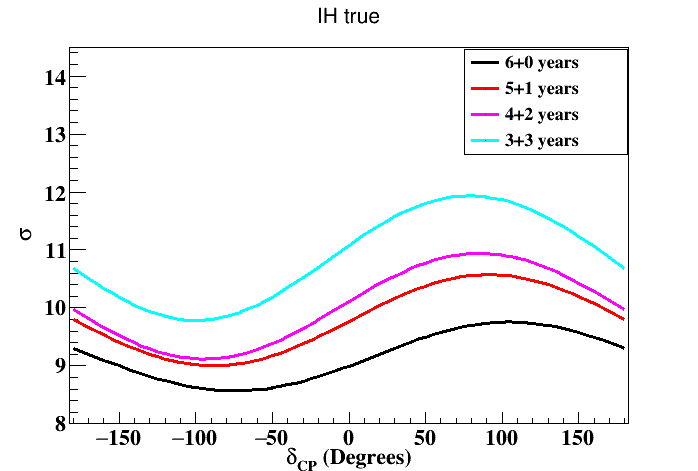}
\caption{\label{mh10_std}  MH sensitivity as a function of true values of $\delta_{cp}$ assuming $10\%$ muon misidentification and for different run times for 6 years. True hierarchy has been assumed to be normal (inverted) for the upper (lower) panel.}
\end{figure}

\begin{figure}[H]
  \centering
 \includegraphics[width=0.9\textwidth]{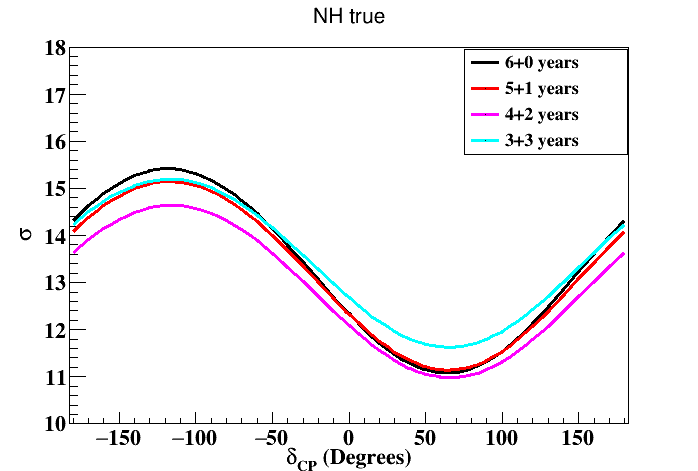}
 \includegraphics[width=0.9\textwidth]{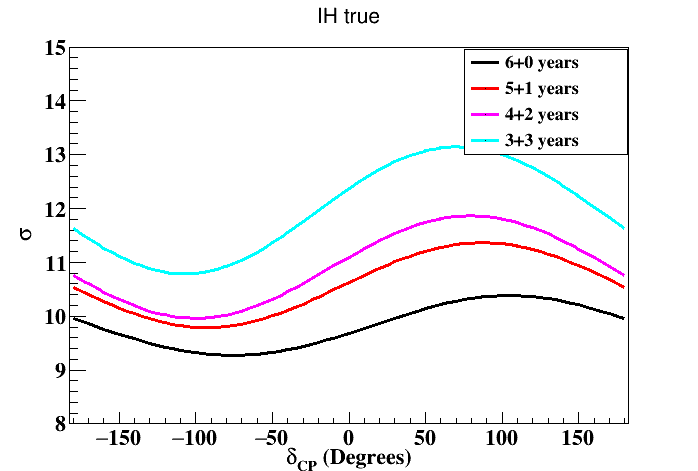}
\caption{\label{mh5_std}  MH sensitivity as a function of true values of $\delta_{cp}$ assuming $5\%$ muon misidentification and for different runtimes. True hierarchy has been assumed to be normal (inverted) for the upper (lower) panel.}
\end{figure}

For simulations, $\chi^2$ and $\dchsq$ are always equivalent.
$\sigma=\sqrt(\chi^{2})$ has been calculated as a function of true values of $\delta_{cp}$ by marginalising it over the test parameter values. The same procedure has been repeated with the considerations that IH is true and NH is test hierarchy. In this study, both disappearance and appearance channels have been considered and combined $\chi^{2}$ for $\nu$ and $\bar{\nu}$ has been shown. MH sensitivity plots have been generated assuming $10\%$ muon mis identification
in P2O detector and are shown in Figure~\ref{mh10_std} and with $5\%$ muon misidentification has been shown in Figure~\ref{mh5_std}.

From fig.~\ref{mh10_std}, it is obvious that for $10\%$ muon mis-identification factor, changes in neutrino and anti-neutrino run time for a total run of 6 years do not affect hierarchy sensitivity when NH is the true hierarchy. However, for IH, changes in run time have significant effect on hierarchy sensitivity and $3+3$ years run has the maximum hierarchy sensitivity. This is because the true value of $\tz$ is in HO, and as we have already shown in section~\ref{prob} with fig.~\ref{prob-uni-oct} that the octant-hierarchy combination HO-IH has the maximum separation from another combination LO-NH for $\pmebar$. Hence, addition of anti-neutrino run improves the hierarchy sensitivity for IH. Another important feature is that for NH $\dcp=-90^\circ$ has the maximum hierarchy sensitivity for the $6+0$ run, where as $\dcp=90^\circ$ has the same for $3+3$ run. This is because $\pme$ for NH and $\dcp=-90^\circ$ is well separated from that for IH for all $\dcp$ values, whereas NH and $\dcp=90^\circ$ has the maximum separation from IH and $\dcp=-90^\circ$ for $\pmebar$. This has already been explained in section~\ref{prob} with fig.~\ref{prob-uni}. It can also be noted that wrong hierarchy can be ruled out for all the $\dcp$ values and all the run times at more than $10\, \sigma$ ($8\, \sigma$) confidence level (C.L.), when NH (IH) is the true hierarchy.

When muon misidentification factor is reduced to $5\%$, mass hierarchy sensitivity of the experiment increases. Now the wrong hierarchy can be ruled out for all $\dcp$ values and all the different run times at more than $11\, \sigma$ ($9\, \sigma$) C.L. when NH (IH) is the true hierarchy. Other features remain same.

\subsection{Mass hierarchy sensitivity with non-unitary oscillation}
\label{hierarchy-nonuni}
\begin{figure}[H]
  \centering
  \includegraphics[width=0.6\textwidth]{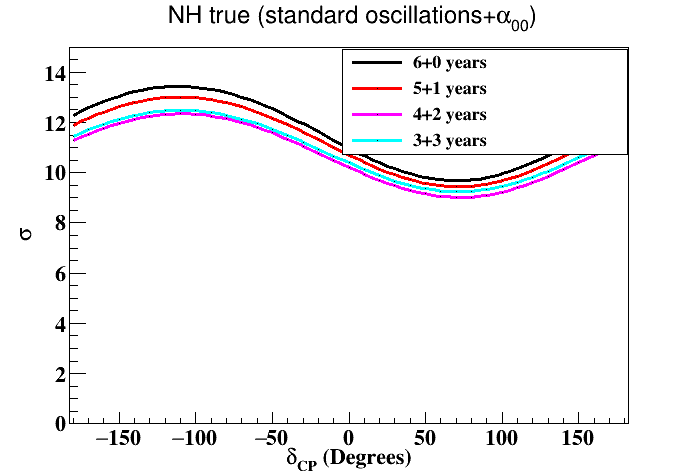}
  \includegraphics[width=0.6\textwidth]{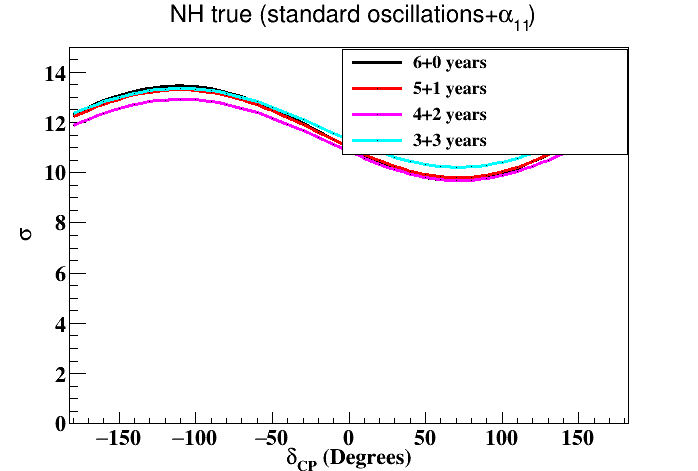}
  \includegraphics[width=0.6\textwidth]{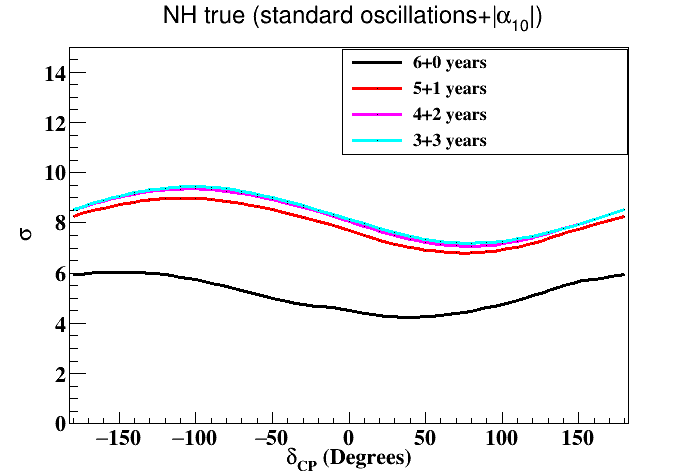}
\caption{\label{mh_nonuni_effnh} Effect of non unitary parameters on MH sensitivity for $ (P2O)$ as a function of true values of $\delta_{cp}$ assuming $10\%$ muon misidentification for various run time years keeping NH as the true and IH as the test hierarchy}
\end{figure}

\begin{figure}[H]
\centering
  \includegraphics[width=0.6\textwidth]{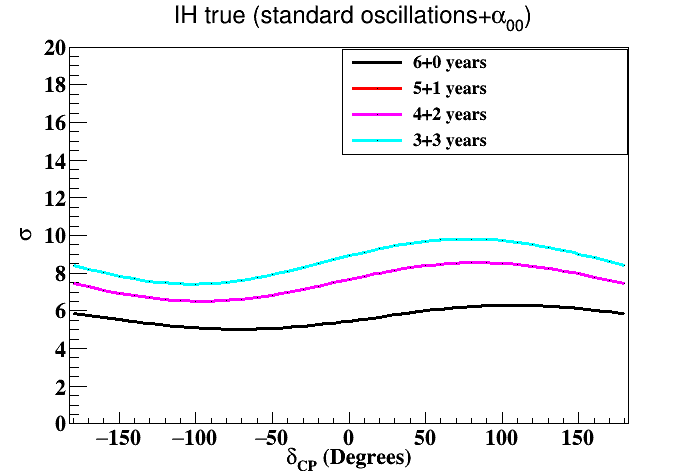}
  \includegraphics[width=0.6\textwidth]{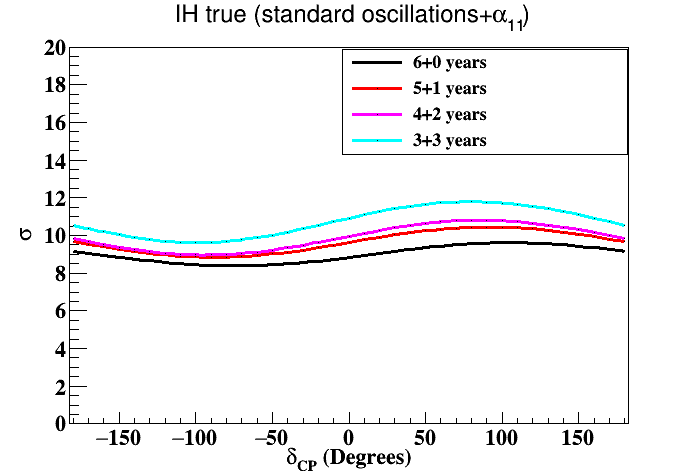}
  \includegraphics[width=0.6\textwidth]{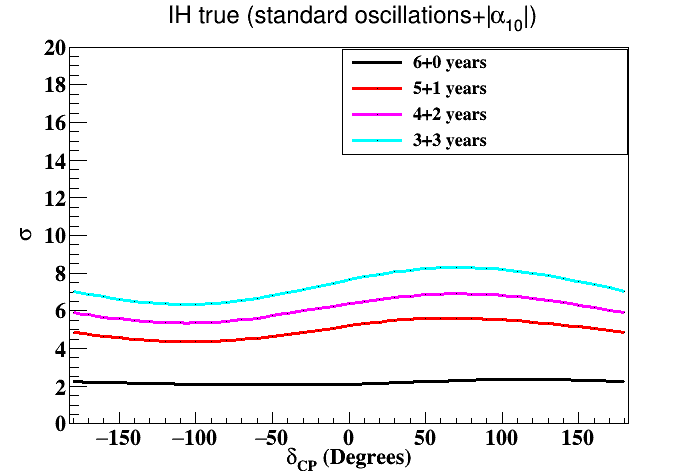}
  
\caption{\label{mh_nonuni_effih} Effect of non unitary parameters on MH sensitivity for P2O as a function of true values of $\delta_{cp}$ assuming $10\%$ muon misidentification for various run time years keeping IH as the true and NH as the test hierarchy}
\end{figure}

In this subsection, the true event rates have been generated with unitary mixing as before. However, here we have assumed non-unitary mixing to calculate the test event rates. We have considered the effect of one non-unitary parameter at a time. For test values of non-unitary parameters, $\alpha_{00}$ and $\alpha_{11}$ have been varied in the range $[0.8:1.0]$, $\alpha_{10}$ has been varied in the range $[0:1.0]$ and $\phi_{10}$ has been kept fixed at $0$. As before, test values of
stanadard oscillation parameters $\sin^2\theta_{23}$, $\dl$ have been varied within their 3$\sigma$ range and  rest of the other oscillation parameters are set at their standard values. Priors on  $\theta_{23}$, $\dl$ have been applied at final $\chi^{2}$ calculation. We have assumed only $10\%$ muon misidentification factor. We have plotted $\sigma$ =$\sqrt(\chi^{2})$ for sensitivity estimation.
We have also shown the effect of  $\alpha_{00}$, $\alpha_{11}$, $\alpha_{10}$  variation on MH sensitivity for various run time years. The effects of individual non-unitary parameters  have been shown for different run times in fig.~\ref{mh_nonuni_effnh} (\ref{mh_nonuni_effih}) when NH (IH) is the true hierarchy.

From fig.~\ref{mh_nonuni_effnh}, we can see that when NH is the true hierarchy, in case of $\alpha_{00}$, and $\alpha_{11}$, mass hierarchy can be established in P2O at more than $10\,\sigma$ C.L. after 6 years. For these two parameters, anti-neutrino run does not have much effect on MH sensitivity. However, for $|\alpha_{10}|$, mass hierarchy can be established at only $6\,\sigma$ C.L. after 6 years of only neutrino run. Anti-neutrino run can improve the sensitivity to $9\,\sigma$.
In case of IH being the true hierarchy (fig.~\ref{mh_nonuni_effih}), in case of $\alpha_{00}$, hierarchy can be determined at $6\, \sigma$, after 6 years of only neutrino run. Anti-neutrino run can improve the sensitivity up to $8\,\sigma$. However, for $\alpha_{11}$, the hierarchy can be established at $\sim 8\,\sigma$ after 6 years, and anti-neutrino run does not have significant effect on hierarchy sensitivity. In case of $|\alpha_{10}|\\$, after 6 years of neutrino run, hierarchy can be established at $2\,\sigma$ C.L. Anti-neutrino run can improve the sensitivity up to $8\,\sigma$.
\section{Establishing CP violation}
\label{CPV}
In this section, we discuss about the CP violation sensitivity, i.e. the sensitivity to rule out CP conserving $\dcp$ values, of P2O, when the data are analysed with both unitary and non-unitary mixing hypothesis. To do so, the true $\dcp$ values have been varied in the complete range. Other standard unitary oscillation parameter values have been fixed at the values given in Table \ref{values}.
\subsection{CP violation sensitivity with standard unitary oscillations}
\label{cpv-uni}
\begin{figure}[H]
   \includegraphics[width=0.9\textwidth]{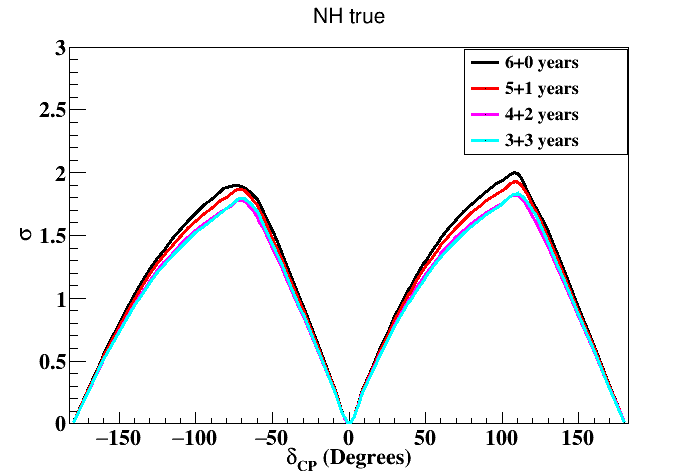}
  \includegraphics[width=0.9\textwidth]{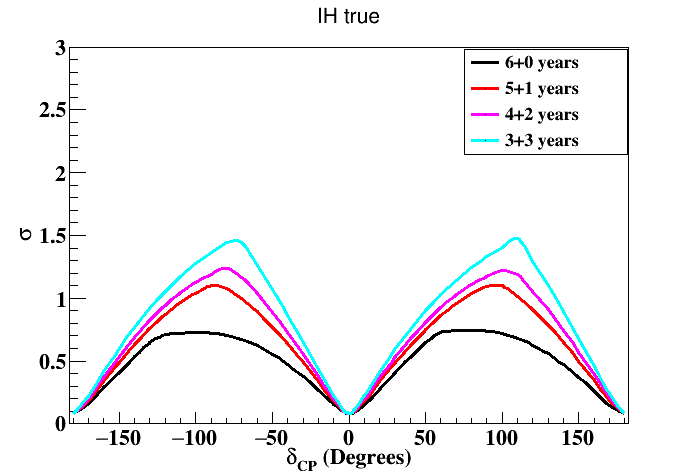}
   \caption{\label{cpv10_std} CPV sensitivity as a function of true values of $\delta_{cp}$ assuming $10\%$ muon misidentification}
\end{figure}

\begin{figure}[H]
  \includegraphics[width=0.9\textwidth]{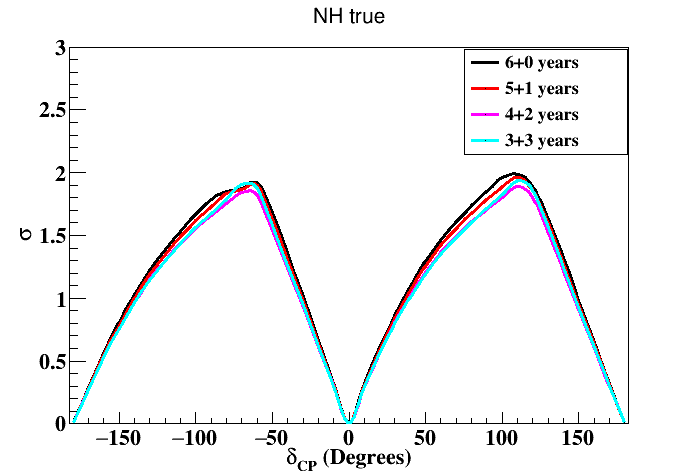}
  \includegraphics[width=0.9\textwidth]{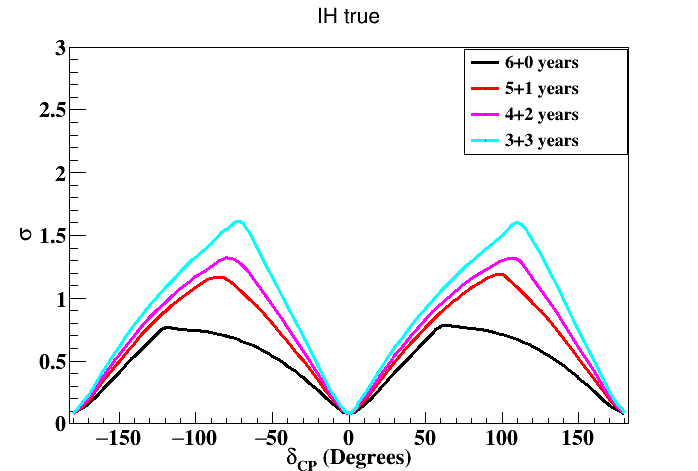}
   \caption{\label{cpv5_std} CP violation sensitivity as a function of true values of $\delta_{cp}$ assuming $5\%$ muon misidentification}
\end{figure}

At first, the test event rates and the $\chi^2$ between true and test event rates have been calculated assuming unitary mixing. The test $\dcp$ values are $-180^\circ$, $0$, and $180^\circ$. Other test parameter values of the unitary oscillation parameters have been varied in the range described in section \ref{hierarchy-uni}. Both NH and IH have been considered as test hierarchy. The $\chi^2$ values have been marginalised over test parameters and in fig.~\ref{cpv10_std}, the CP violation sensitivity has been shown as a function of true $\dcp$ values for $10\%$ muon mis-identification factor. For NH true, addition of anti-neutrino run does not have any significant effect to CP violation sensitivity. However, for IH true, increasing anti-neutrino run improves the CP violation sensitivity with maximum sensitivity achieved at 3 years of neutrino and anti-neutrino run each. For both the hierarchies, Maximum CP violation sensitivity of $1.5\, \sigma$ can be achieved for $\dcp=\pm 90^\circ$. When the muon mis-identification factor is reduced to $5\%$, the maximum CP violation sensitivity after 6 years can be $2\,\sigma$ for $\dcp=\pm 90^\circ$. This has been shown in fig.~\ref{cpv5_std}.
\subsection {CP violation sensitivity with non-unitary oscillation}
\label{cpv-non-uni}
\begin{figure}[H]
\centering
  \includegraphics[width=0.65\textwidth]{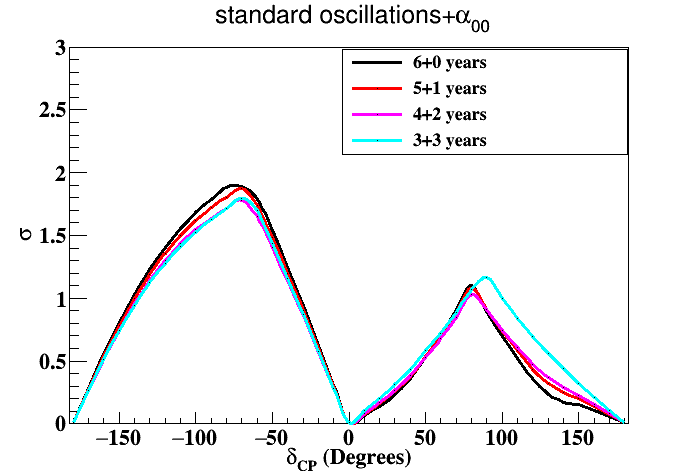}
  \includegraphics[width=0.65\textwidth]{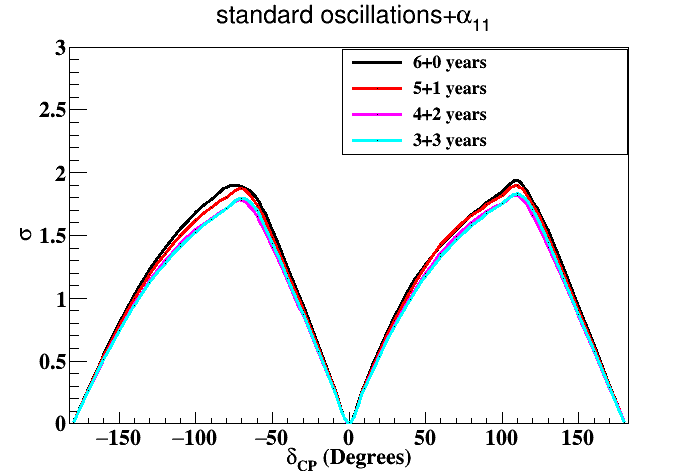}
  \includegraphics[width=0.65\textwidth]{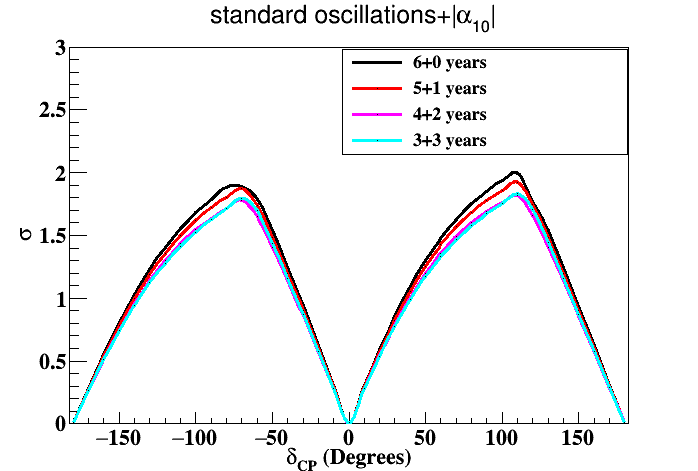}
\caption{\label{cpv_nonuni_nh} CP violation sensitivity including non unitary parameters for$ (P2O)$as a function of true values of $\delta_{cp}$ assuming $10\%$ muon misidentification for various run time years keepin NH is true hierarchy}
\end{figure}

\begin{figure}[H]
\centering
  \includegraphics[width=0.65\textwidth]{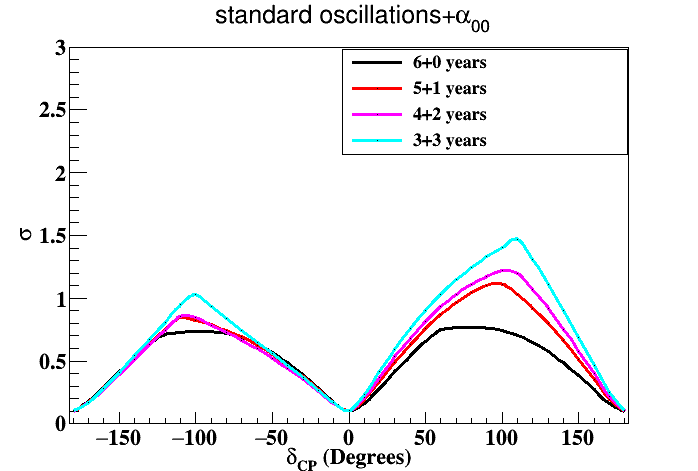}
  \includegraphics[width=0.65\textwidth]{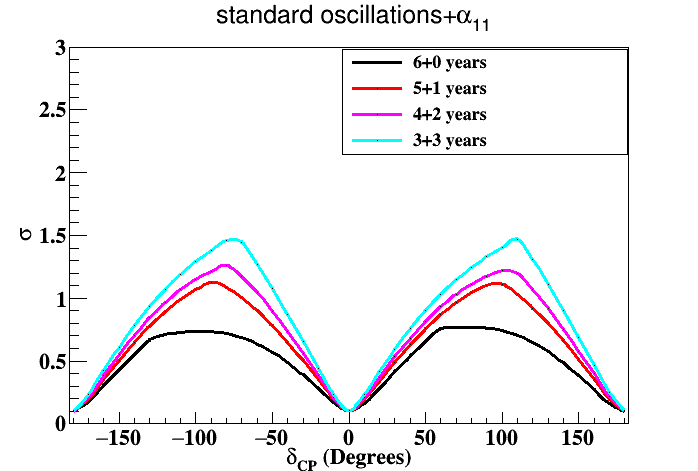}
  \includegraphics[width=0.65\textwidth]{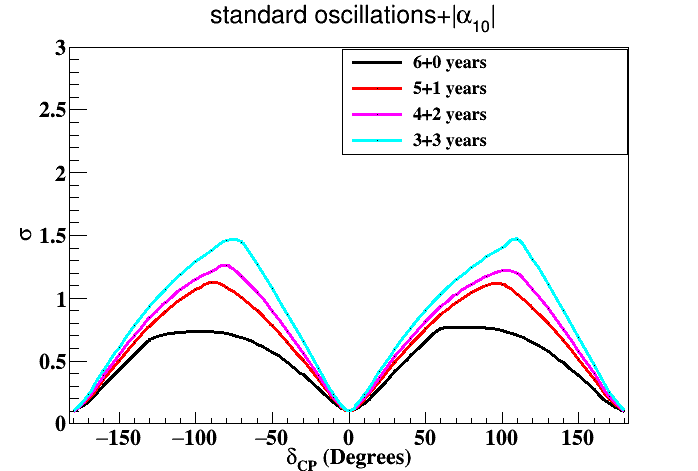}
\caption{\label{cpv_nonuni_ih} CP violation sensitivity including non unitary parameters for$ (P2O)$as a function of true values of $\delta_{cp}$ assuming $10\%$ muon misidentification for various run time years keepin IH is true hierarchy}
\end{figure}

Now the test event rates have been calculated assuming non-unitary $3\times 3$ neutrino mixing. Test $\dcp$ values are $-180^\circ$, $0$, and $180^\circ$. Other standard oscillation parameter ranges are same as described in section \ref{hierarchy-uni}. We have observed the effect of one non-unitary parameter at a time on the CP violation sensitivity. The non-unitary parameters have been varied in the range described in section \ref{hierarchy-nonuni}. Both NH and IH have been considered as the true hierarchy. The $\chi^2$ values have been marginalised over test parameter values. Effects of individual non-unitary parameters on the CP violation sensitivity has been shown in fig.~\ref{cpv_nonuni_nh} and fig.~\ref{cpv_nonuni_ih} for NH and IH being the true hierarchy respectively. Muon mis-identification factor is $10\%$. 
From fig.~\ref{cpv_nonuni_nh}, we can see that anti-neutrino run does not have any significant effect on CP violation sensitivity at P2O in the case of non-unitary mixing when NH is the true hierarchy. In case of $\alpha_{11}$, and $|\alpha_{10}|$, CP violation can be established at more than $1.5\, \sigma$ C.L. for true $\dcp=\pm 90^\circ$. However, for $\alpha_{00}$, CP violation can be established only at $1\,\sigma$ ($1.5\,\sigma$) for true $\dcp=90^\circ$ ($-90^\circ$).

In case of IH being the true hierarchy, CP violation can be established only at $0.5\, \sigma$ for $|\alpha_{10}|$, and $\alpha_{11}$ after 6 years of neutrino run when $\dcp=\pm 90^\circ$. Anti-neutrino run can improve the sensitivity up to $\sim 1.5\, \sigma$ C.L. For $\alpha_{00}$ also, CPV can be established only at $0.5\, \sigma$ after 6 years of neutrino run when $\dcp=\pm 90^\circ$. However, anti-neutrino run can improve it up to $1\, \sigma$ ($1.5\, \sigma$) for $\dcp=-90^\circ$ ($90^\circ$).
\section{Constraint on non unitary parameters}
\label{const}
\begin{figure}[H]
  \centering
  \includegraphics[width=0.49\textwidth]{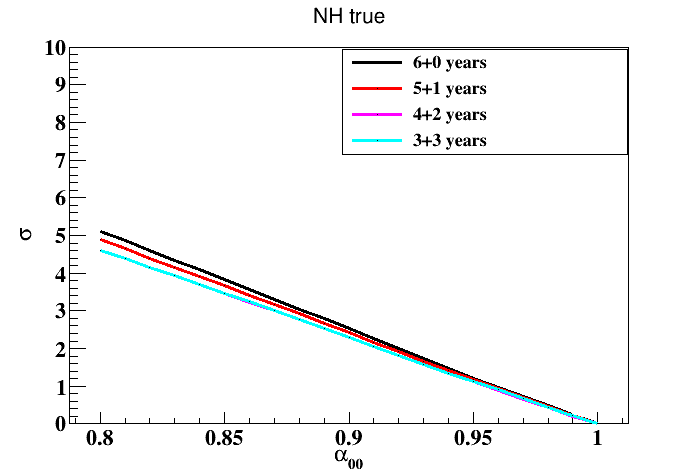}
  \includegraphics[width=0.49\textwidth]{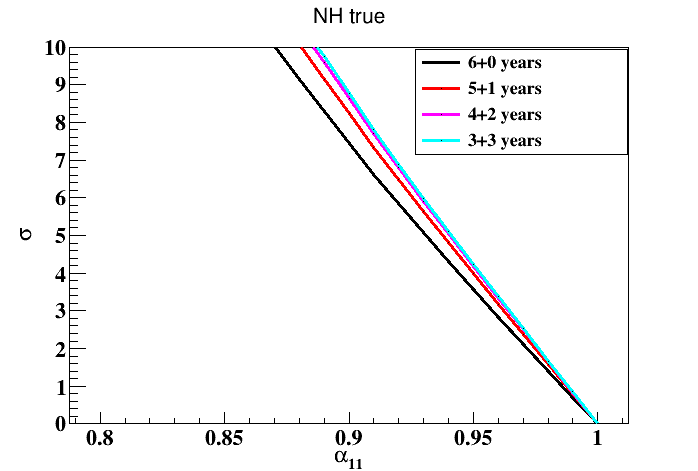}
  \includegraphics[width=0.49\textwidth]{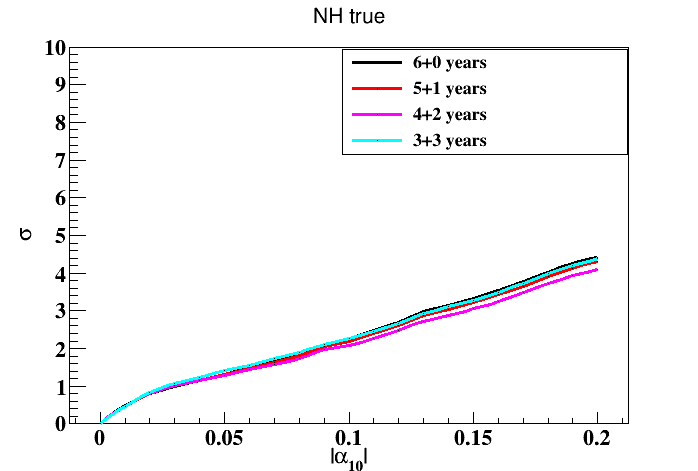}
\caption{\label{const_nh_nh} $\sigma=\sqrt{\dchsq}$ as a function of individual non-unitary parameters for NH being the true and test hierarchies and for different run times.}
\end{figure}

\begin{figure}[H]
  \centering
  \includegraphics[width=0.49\textwidth]{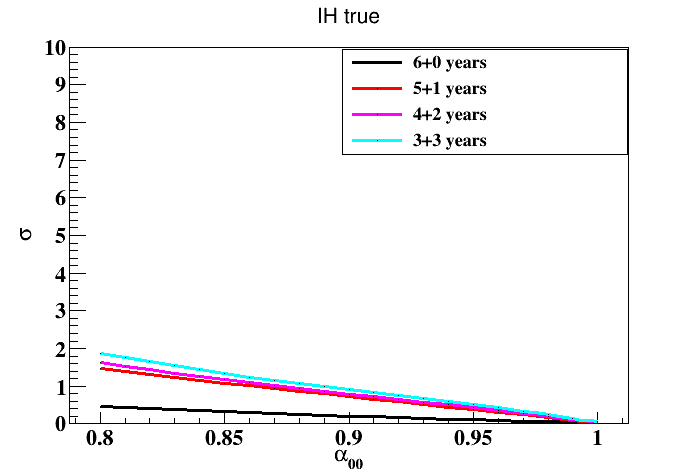}
  \includegraphics[width=0.49\textwidth]{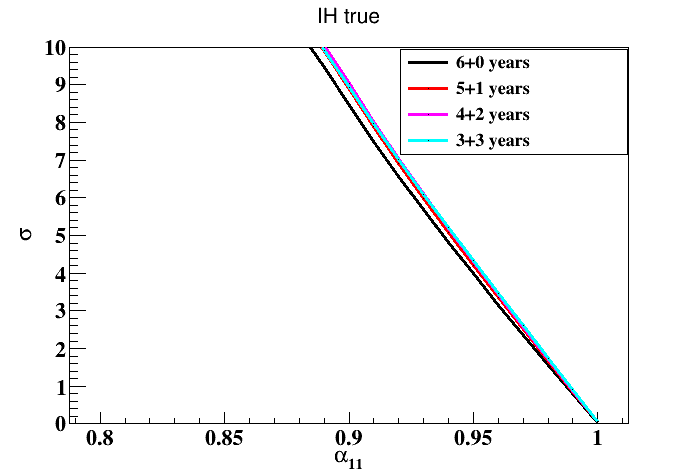}
  \includegraphics[width=0.49\textwidth]{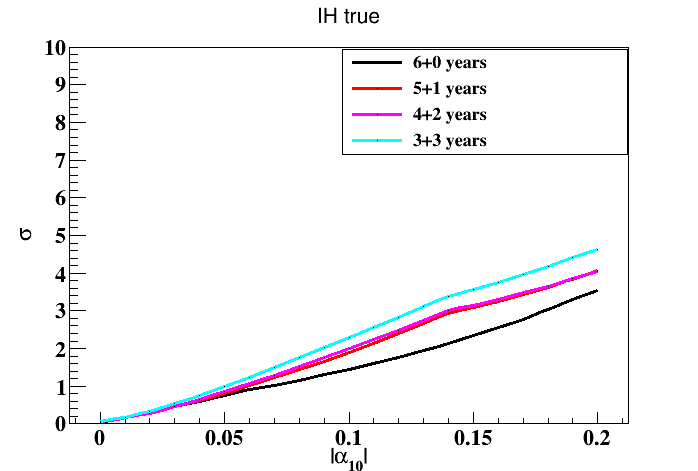}
\caption{\label{const_ih_ih} $\sigma=\sqrt{\dchsq}$ as a function of individual non-unitary parameters for IH being the true and test hierarchies and for different run times.}
\end{figure}

In this section, we have tried to derive the constraints on the non-unitary parameters that P2O can put on them in the future. To do so, we calculated the true event rates with the oscillation parameter values fixed at their present global best-fit values \cite{nufit, Esteban:2020cvm}. We varied the test values of $\dcp$ in the range $[-180^\circ:180^\circ]$ and $\sin^2 \tz$ in its $3\, \sigma$ range. The non-unitary parameter, on which we want to derive the constraint was varied, and other non-unitary parameters were fixed at their unitary values. $\phi_{10}$ has been fixed at $0$. In fig. \ref{const_nh_nh} we have shown $\dchsq$ as a function of the non-unitary parameters for NH being both the true and test hierarchy. We can see that P2O can provide constraint at $90\%$ ($3\, \sigma$) confidence level for $\alpha_{00}> 0.92$ (0.87), $\alpha_{11}>0.98$ (0.96), and $|\alpha_{10}|<0.06$ (0.13) after $3+3$ years when both true and test hierarchies are NH. In fig.~\ref{const_ih_ih}, we have shown the similar results for IH being the true and test hierarchy. It can be seen that the constraint at $90\%$ ($3\,\sigma$) confidence level after $3+3$ years are $\alpha_{00}> 0.82$ (out of range), $\alpha_{11}>0.98$ (0.96), and $|\alpha_{10}|<0.07$ (0.12). These constraints are less stringent than the present constraints from the global analysis \cite{Escrihuela:2016ube}, but better than the constraints put on by the present \nova and T2K experimental data \cite{Miranda:2019ynh}.

\section{Conclusion}
\label{conclusion}
We found out that in case of unitary mixing, the hierarchy-$\dcp$, and octant-hierarchy degeneracies present at experiments like \nova and T2K breaks down for P2O. Now, the wrong hierarchy can be ruled out at more that $9\, \sigma$ ($8\, \sigma$) C.L. after 6 years for all the $\dcp$ values when NH (IH) is the true hierarchy and the muon mis-identification factor is $10\%$. Addition of anti-neutrino run does not have any significant effect on hierarchy sensitivity when NH is true. However, the hierarchy sensitivity significantly improves after addition of anti-neutrino run and for 3 years each of equal neutrino and anti-neutrino run, NH can be ruled out at more than $10\,\sigma$ C.L when IH is the true hierarchy. When the muon mis-identification factor is reduced to $5\,\%$, the hierarchy sensitivity is improved by around $1\,\sigma$. 

CP violation can be established at $2\, \sigma$ (more than $0.5\, \sigma$) for $\dcp=\pm 90^\circ$ for NH (IH) being the true hierarchy when the muon mis-identification factor is $10\%$. Just like the case of mass hierarchy sensitivity, addition of anti-neutrino run does not have any significant effect when NH is the true hierarchy, but has a significant effect for IH improving the CP violation sensitivity to $1.5\, \sigma$ at $\dcp=\pm 90^\circ$ for equal neutrino and anti-neutrino runs of 3 years each. When muon mis-identification factor is improved to $5\%$, the CP violation sensitivity is only slightly improved. 

We have calculated the hierarchy and CP violation sensitivity assuming $10\%$ muon mis-identification in case of non-unitary mixing by observing effect of one parameter at a time. When NH is the true hierarchy, the wrong hierarchy can be excluded at more than $10\, \sigma$ C.L. after 6 years for both $\alpha_{00}$, and $\alpha_{11}$. Anti-neutrino run does not have much effect in these two cases. For $|\alpha_{10}|$, hierarchy can be established at $\sim 6\, \sigma$ after 6 years of only neutrino run when NH is the true hierarchy.Anti-neutrino run can improbe it up to $8\,\sigma$. When IH is the true hierarchy, the wrong hierarchy can be excluded at $6\,\sigma$ (more than $8\,\sigma$) after 6 years of only neutrino run for $\alpha_{00}$ ($\alpha_{11}$). Anti-neutrino run can improve the sensitivity up to $8\,\sigma$ ($10\,\sigma$) for $\alpha_{00}$ ($\alpha_{11}$). In case of $|\alpha_{10}|$, hierarchy can be established only at $2\, \sigma$ after only 6 years of neutrino run when IH is true. It can be improved up to $8\, \sigma$ after anti-neutrino run.

For non-unitary mixing, anti-neutrino run does not have any significant effect on CP violation sensitivity when NH is the true hierarchy. In case of $\alpha_{11}$, and $|\alpha_{10}|$, CP violation can be established at more than $1.5\,\sigma$ for $\dcp=\pm90^\circ$. For $\alpha_{00}$, CP violation can be established at more than $1.5\, \sigma$ ($1\, \sigma$) for $\dcp=90^\circ$ ($-90^\circ$). When IH is the true hierarchy, anti-neutrino run can improve CP violation sensitivity significantly. For all three non-unitary parameters: $\alpha_{00}$, $\alpha_{11}$, and $|\alpha_{10}|$, CP violation can be established at $\sim 0.5\, \sigma$ for $\dcp=\pm 90^\circ$. In case of $\alpha_{00}$, anti-neutrino run can be improved up to $1\,\sigma$ ($1.5\, \sigma$) when $\dcp=-90^\circ$ ($90^\circ$). In cases of $\alpha_{11}$, and $|\alpha_{10}|$, CP violation can be improved by anti-neutrino run up to $1.5\,\sigma$ for $\dcp=\pm 90^\circ$ when IH is the true hierarchy.

We have also calculated the constraints that can be put on $\alpha_{00}$, $|\alpha_{11}|$, and $|\alpha_{10}|$ by the P2O in the future. We found out that the constrained at $90\%$ ($3\, \sigma$) confidence level can be achieved for $\alpha_{00}> 0.92$ (0.87), $\alpha_{11}>0.98$ (0.96), and $|\alpha_{10}|<0.06$ (0.13) after 3 years each of neutrino and anti-neutrino run when both true and test hierarchies are NH. For IH, the future possible constraints at $90\%$ ($3\,\sigma$) confidence level are $\alpha_{00}> 0.82$ (out of range), $\alpha_{11}>0.98$ ($0.96$), and $|\alpha_{10}|<0.07$ ($0.12$) after 3 years of neutrino and 3 years of anti-neutrino run.

\section*{Acknowledgement} 
U.R. was supported by a grant from the University of Johannesburg Research Council.

\bibliographystyle{apsrev}

 \bibliography{referenceslist}
\end{document}